%% file: RL_paper.tex
\documentclass[a4paper,11pt]{article}
\pdfoutput=1 

\usepackage{jheppub} 
                     
\usepackage{amsmath}

\usepackage{color}

\usepackage[switch, modulo]{lineno}

\title{\boldmath Boosting background suppression in the NEXT experiment through Richardson-Lucy deconvolution}

\input{src/Authors3.tex}

\abstract{
	Next-generation neutrinoless double beta decay experiments aim for half-life sensitivities of $\sim10^{27}$ yr, requiring suppressing backgrounds to $<1$ count/tonne/yr. For this, any extra background rejection handle, beyond excellent energy resolution and the use of extremely radiopure materials, is of utmost importance. The NEXT experiment exploits differences in the spatial ionization patterns of double beta decay and single-electron events to discriminate signal from background. While the former display two Bragg peak dense ionization regions at the opposite ends of the track, the latter typically have only one such feature. Thus, comparing the energies at the track extremes provides an additional rejection tool. The unique combination of the topology-based background discrimination and excellent energy resolution (1\% FWHM at the Q-value of the decay) is the distinguishing feature of NEXT. Previous studies demonstrated a topological background rejection factor of $\sim5$ when reconstructing electron-positron pairs in the $^{208}$Tl 1.6 MeV double escape peak (with Compton events as background), recorded in the NEXT-White demonstrator at the Laboratorio Subterr\'aneo de Canfranc, with 72\% signal efficiency. This was recently improved through the use of a deep convolutional neural network to yield a background rejection factor of $\sim10$ with 65\% signal efficiency. Here, we present a new reconstruction method, based on the Richardson-Lucy deconvolution algorithm, which allows reversing the blurring induced by electron diffusion and electroluminescence light production in the NEXT TPC. The new method yields highly refined 3D images of reconstructed events, and, as a result, significantly improves the topological background discrimination. When applied to real-data 1.6 MeV $e^-e^+$ pairs, it leads to a background rejection factor of 27 at 57\% signal efficiency.
} 


\begin{document}
\maketitle
\flushbottom

\input{src/intro.tex}

\input{src/NEXT-White.tex}

\input{src/blurring.tex}
\input{src/RL3.tex}

\input{src/pp_analysis.tex}

\input{src/discussion.tex}

\acknowledgments
The NEXT Collaboration acknowledges support from the following agencies and institutions: the European Research Council (ERC) under the Advanced Grant 339787-NEXT; the European Union's Framework Programme for Research and Innovation Horizon 2020 (2014--2020) under the Grant Agreements No.\ 674896, 690575 and 740055; the Ministerio de Econom\'ia y Competitividad and the Ministerio de Ciencia, Innovaci\'on y Universidades of Spain under grants FIS2014-53371-C04, RTI2018-095979, the Severo Ochoa Program grants SEV-2014-0398 and CEX2018-000867-S, and the Mar\'ia de Maeztu Program MDM-2016-0692; the Generalitat Valenciana under grants PROMETEO/2016/120 and SEJI/2017/011; the Portuguese FCT under project PTDC/FIS-NUC/2525/2014 and under projects UID/04559/2020 to fund the activities of LIBPhys-UC; the U.S.\ Department of Energy under contracts No.\ DE-AC02-06CH11357 (Argonne National Laboratory), DE-AC02-07CH11359 (Fermi National Accelerator Laboratory), DE-FG02-13ER42020 (Texas A\&M) and DE-SC0019223 / DE-SC0019054 (University of Texas at Arlington); the University of Texas at Arlington (USA); and the Pazy Foundation (Israel) under grants 877040 and 877041. DGD acknowledges Ramon y Cajal program (Spain) under contract number RYC-2015-18820. JM-A acknowledges support from Fundaci\'on Bancaria ``la Caixa'' (ID 100010434), grant code LCF/BQ/PI19/11690012. AS acknowledges support from the Kreitman School of Advanced Graduate Studies at Ben-Gurion University.

\input{src/RL_app.tex}

\bibliographystyle{JHEP}
\bibliography{biblio}

\end{document}

%% file: src/Authors3.tex
\collaboration{The NEXT Collaboration}
\author[6,a]{A.~Sim\'on, \note[a]{Corresponding author.}}
\author[6,7]{Y.~Ifergan,}
\author[6]{A.B.~Redwine,}
\author[6,b]{R.~Weiss-Babai,\note[b]{Now in Soreq Nuclear Research Center, Yavneh, Israel.}}
\author[6,a]{L.~Arazi,}
\author[20]{C.~Adams,}
\author[17]{H.~Almaz\'an,}
\author[25]{V.~\'Alvarez,}
\author[18]{B.~Aparicio,}
\author[18]{A.I.~Aranburu,}
% ...
\emailAdd{ander@post.bgu.ac.il}
\emailAdd{larazi@post.bgu.ac.il}
% ...
\affiliation[6]{
Unit of Nuclear Engineering, Faculty of Engineering Sciences, Ben-Gurion University of the Negev, Beer-Sheva, Israel}
\affiliation[7]{
Nuclear Research Center Negev, Beer-Sheva, Israel}
% ...
\affiliation[17]{
Donostia International Physics Center (DIPC), Donostia-San Sebasti\'an, Spain}
\affiliation[18]{
Faculty of Chemistry, University of the Basque Country (UPV/EHU), Manuel de Lardizabal 3, 20018 Donostia-San Sebasti\'an, Spain}
% ...
\affiliation[20]{
Argonne National Laboratory, Lemont, Illinois, USA}
% ...
\affiliation[25]{
Instituto de Instrumentaci\'on para Imagen Molecular (I3M), CSIC \& Univ.\ Polit\`ecnica de Val\`encia, Valencia, Spain}
% ...

%% file: src/intro.tex
\section{Introduction}
\label{sec:next}

The search for neutrinoless double beta ($0\nu\beta\beta$) decay is the most promising experimental path to determine whether the neutrino is a Majorana fermion, with far-reaching implications in particle physics and cosmology \cite{Avignone2008, Davidson2008, Blennow2010, Dell?Oro2016, Dolinski2019}. Presently, several collaborations pursue different technologies for detecting $0\nu\beta\beta$ decay with the leading experiments focusing on $^{76}$Ge \cite{GERDA2020, Majorana2019, LEGEND2017}, $^{136}$Xe \cite{KamLAND-Zen2016, Loaded_scintillators2019, EXO2018, nEXO2018, DARWIN2020, JJ2019, PandaX-III}, $^{130}$Te \cite{CUORE2020, SNO+2016, Loaded_scintillators2019}, and $^{100}$Mo \cite{CUPID2019, CUPID-Mo_results, AMORE2019}. The long half-life of $0\nu\beta\beta$ decay (above 1.8 $\times$ 10$^{26}$ yr in $^{76}$Ge \cite{GERDA2020} and 1.07 $\times$ 10$^{26}$ yr in $^{136}$Xe \cite{KamLAND-Zen2016}) makes its detection extremely difficult, with only a few candidate $0\nu\beta\beta$ events expected throughout the running life of an experiment, calling for outstanding background suppression capabilities. The next generation of $0\nu\beta\beta$ decay experiments will aim at half-life sensitivities of $\sim10^{27}$\;yr, requiring, in turn, a background level below $\sim1$ counts/tonne/yr.

\begin{figure}
	\centering
	\includegraphics[trim={0 0 0 0}, clip, height=6cm]{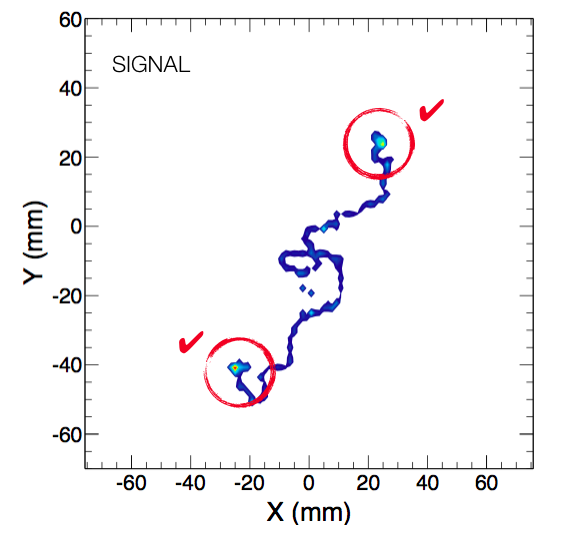}
	\includegraphics[trim={0 0 0 0}, clip, height=6cm]{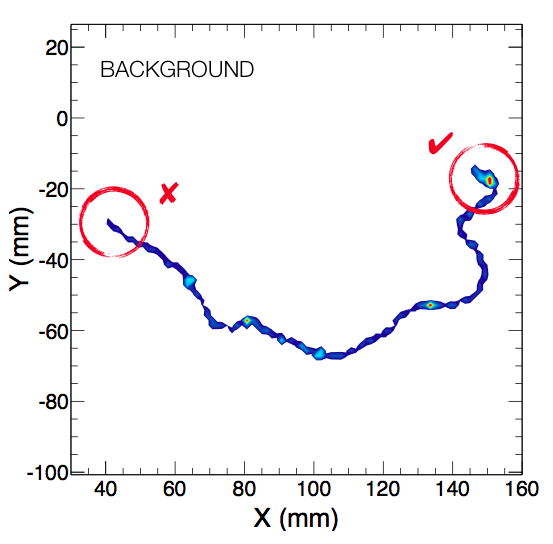}
	\caption{Simulated GEANT4 signal and background tracks near $Q_{\beta\beta}$ at 15 bar, from \cite{Ferrario:2015kta}.}
	\label{img:geant4_tracks}
\end{figure}

NEXT (Neutrino Experiment with a Xenon TPC) is a staged experimental program aiming at the detection of $0\nu\beta\beta$ decay in $^{136}$Xe, using successive generations of high-pressure gaseous xenon electroluminescent time projection chambers (HPXe EL-TPC) \cite{Nygren:2009zz}. The choice of gaseous rather than liquid xenon is driven by two considerations: (1) the attainable energy resolution at the Q-value of the decay, $Q_{\beta\beta}$ (which for $^{136}$Xe is 2458\;keV), is a factor 3 better in gas than in liquid \cite{Bolotnikov:1997, Renner:2019pfe, EXO2018, XENON1T_Eres}; and (2) whereas in liquid xenon events are point-like, the projected length of ionization tracks in high-pressure Xe gas is $\sim10$\;cm, allowing for background discrimination based on the track topology, as first pioneered by the Gotthard experiment \cite{Gotthard1998}. The distinguishing topological feature of a two-electron double beta decay event in high-pressure gas is the appearance of two blob-like Bragg-peak energy depositions at the opposite ends of the track, in contrast with single-electron background events which have only one such feature (figure \ref{img:geant4_tracks}). Thus, by reconstructing the track in 3D and comparing the energy contained in small regions at its extremities, one can effectively classify the event as either signal or background. 

The current stage of the NEXT program is the NEXT-White\footnote{Named after Prof. James White, our late mentor and friend.} demonstrator, whose TPC has an active volume half a meter in diameter and length \cite{Monrabal:2018xlr}. NEXT-White (``NEW'') is a radiopure detector, operated underground under low-background conditions at the Laboratorio Subterr\'aneo de Canfranc (LSC), using xenon enriched to $90\%$ $^{136}$Xe. Its purpose is to  validate all aspects of the technology on a large scale, including full characterization of the background model and the technique's background rejection power, and to demonstrate its performance on two-neutrino double beta ($2\nu\beta\beta$) decay events. NEXT-White, which has been running continuously since October 2017, will be superseded in 2022 by the twice-larger NEXT-100 detector, which will deploy 97\;kg of enriched xenon and demonstrate sensitivity to $0\nu\beta\beta$ decay half-lives on the scale of $10^{26}$\;yr \cite{JJ2019}. This will pave the way to a tonne-scale experiment which will be sensitive to half-lives longer than $10^{27}$\;yr \cite{NEXT_tonne}. Importantly, the NEXT Collaboration pursues in parallel an extensive R\&D program to develop the capability of detecting the $^{136}$Ba daughter resulting from $^{136}$Xe double beta decays inside a running TPC using {\it single molecule fluorescence imaging} \cite{Jones2016, McDonnald_PRL, Byrnes_2019, Thapa2019, Rivilla2020}. If successful, these efforts could boost the sensitivity of HPXe EL-TPCs to half-lives on the scale of $10^{28}$\;yr.    

NEXT excellent energy resolution -- demonstrated to be $\sim1\%$ full-width half maximum (FWHM) at $Q_{\beta\beta}$ in NEXT-White \cite{Renner:2019pfe} -- 
allows to reduce background from natural radioactivity and $2\nu\beta\beta$ events in the $Q_{\beta\beta}$ region-of-interest (ROI) by about 5 and 12 orders of magnitude, respectively, making the latter completely negligible \cite{NEXT100_sensitivity}. NEXT also utilizes track multiplicity, i.e., the number of distinct tracks in a given event, to reduce background from $\gamma$-ray interactions or from single beta decays with energy in the $Q_{\beta \beta}$ ROI (from cosmogenically-produced $^{137}$Xe). This effectively removes multiple Compton scatters, as well as fast-electron tracks with accompanying vertices of bremsstrahlung and characteristic x-ray interactions, resulting in an additional background rejection factor of $\sim10$ while retaining $\sim70\%$ of the signal \cite{NEXT100_sensitivity, NEXT_tonne}. Finally, using the topology of the remaining single tracks, background is further suppressed by measuring the energy deposition at the track ends as described above.

Tests of the topological background rejection in NEXT-White rely on the use of a $^{228}$Th calibration source to produce 2615\;keV $\gamma$-rays by the decay of $^{208}$Tl. Interactions of these $\gamma$-rays inside the gas lead to the production of electron-positron pairs with a total kinetic energy of 1593\;keV. These leave a trace with two Bragg-peak energy depositions at the track endpoints, mimicking double beta decay events. 
Previous topological analysis of such events in NEXT-White (with Compton electrons serving as background) yielded a background rejection factor of $\sim5$, while retaining 72\% of the signal events \cite{Ferrario:2019kwg} for optimal performance. Extended by Monte Carlo to $Q_{\beta\beta}$, this analysis provided a background rejection factor 7.4 with similar signal efficiency. (Note that these values correspond to the performance of the topological analysis alone, after the preceding energy and single-track cuts.) Recently, these results were improved by using a deep convolutional neural network yielding a background rejection factor of $\sim10$ with a signal acceptance of $\sim$65\% \cite{Kekic:2020cne}.

In this work we describe an improved methodology for track reconstruction. The underlying idea is to enhance the sharpness of reconstructed tracks -- degraded by electron diffusion and spread of light produced in the electroluminescence (EL) process -- through the application of an image deblurring procedure, namely the Richardson-Lucy (RL) deconvolution algorithm \cite{Richardson:72, Lucy:1974yx}. We begin by describing signal production and event reconstruction in NEXT-White, with particular emphasis on previous work regarding the topological analysis. This is followed by a discussion of the blurring effects of diffusion and EL light production, and their quantification through a spatially-dependent point spread function (PSF). We then describe the implementation of the RL algorithm, which is outlined in appendix\;\ref{sec:RL_app}, within the experiment's data processing chain, and validate it by demonstrating the accurate reconstruction of point-like events (individual and pairs) and muon tracks. The procedure is subsequently employed on 1.6\;MeV $e^-e^+$ pairs and gamma-induced background events in both Monte Carlo and data recorded in NEXT-White, yielding a major improvement in topological background rejection. We discuss the effect of the key parameters of the deconvolution procedure on its performance, and present the optimized results.

%% file: src/NEXT-White.tex
\section{NEXT-White: event reconstruction and prior work on topological analysis}
\label{sec:NEXT-White}

The NEXT-White TPC \cite{Monrabal:2018xlr} consists of a cathode and gate grids defining a 53\;cm-long drift region, a transparent anode plate positioned 6 mm behind the gate, and a field cage with an inner diameter of 45\;cm. An array of 12 Hamamatsu R11410-10 3'' photomultiplier tubes (PMTs), constituting the {\it energy plane}, is located 13\;cm behind the cathode. At the opposite end of the TPC, 2\;mm behind the anode plate, an array of 1792 SensL series-C 1\;mm$^2$ silicon photomultipliers (SiPMs) distributed at a pitch of 10 mm, serves as the {\it tracking plane}. The entire tracking plane area is covered by polytetrafluoroethylene (PTFE) to reflect light towards the PMTs, with holes for the SiPMs. The interior wall of the field cage comprises a PTFE tube coated with a thin wavelength-shifting layer of tetraphenyl butadiene (TPB) to further optimize light collection. The anode is a 3 mm-thick fused silica plate, coated on both faces with transparent resistive and conductive layers, which are themselves coated by TPB. 
The detector is presently operated under voltages which define a uniform drift field of 0.42\;kV/cm between the cathode and gate, and a nominal electroluminescence (EL) field of 12.8\;kV/cm between the gate and anode (the {\it EL gap}).  

When an event occurs inside the sensitive volume, the associated primary charged particles form Xe excimers and electron-ion pairs along their track. De-excitation of the former produces prompt vacuum ultra violet (VUV) scintillation light (``S1'') centered at 172\;nm, lasting a few hundred ns. This light, which is shifted to $\sim$430\;nm by the TPB coating the inner surfaces of the TPC and recorded by the PMTs with a sampling time of 25\;ns, provides the start time $t_0$ of the event. The drift field prevents electron-ion recombination and drives the electrons at a uniform velocity of $0.91$\;mm/$\mu$s (for $E_{drift}=0.42$\;kV/cm) towards the EL gap. As they cross it (in $\sim1.6\;\mu$s), they produce EL light (``S2''), which is also centered at 172\;nm and wavelength-shifted by TPB. For a nominal EL field of 12.8\;kV/cm and 6\;mm gap, this provides, at 10\;bar, a light yield of $\sim450$ VUV photons per electron crossing the gap \cite{Freitas2010}. Depending on the track length, the duration of the S2 signal varies from a few $\mu$s to a few hundred $\mu$s. The S2 light is recorded by both the PMTs and SiPMs, with the latter integrated in 1\;$\mu$s time slices. 

During operation, NEXT-White is continuously calibrated using 
$^{83m}$Kr, which is introduced into the gas system by an in-line $^{83}$Rb source. $^{83m}$Kr spreads uniformly throughout the TPC volume and produces point-like energy depositions of 41.6\;keV. These events serve for precise determination of the electron lifetime and $xy$ S2 response of the detector, to generate 3D correction maps which are then employed for accurate energy measurements of signal and background events \cite{Martinez-Lema:2018ibw}. Furthermore, as detailed below, $^{83m}$Kr can also be used to obtain the point-spread function that describes both electron diffusion and the optical response of the tracking plane, a vital ingredient for image deconvolution.

When processing events with energy higher than 400\;keV, SiPM waveforms are re-binned into 2\;$\mu$s, keeping only sensors recording more than 5 photoelectrons (PEs) within each slice. At this stage a second, optional, higher charge threshold is applied. For each SiPM passing this threshold in each slice, a 3D ``hit'' is generated, whose $xy$ coordinates are those of the associated SiPM, the $z$ coordinate is the product of the electron drift velocity and time difference between the slice and S1, and the magnitude is the number of detected PEs. The energy measured by the PMTs in the same time slice is divided among the reconstructed hits, proportionally to the charge of their respective SiPMs. Afterwards the energy assigned to each hit is multiplied by the correction factors derived from the $^{83m}$Kr 3D maps. Figure 2 (left) shows an example of a 3D hit map of a reconstructed track with color representing the number of PEs detected. 

The resulting hits are grouped in cubic voxels with an event-dependent fixed size. These are further grouped into connected tracks, using the {\it breadth-first search} (BFS) algorithm \cite{Cormen2001_intro_algorithms, Ferrario:2015kta}. Voxels are considered to be part of the same track if they have a common face, side or corner. An event may contain more than one track (for example, in the case of multiple Compton scatters). After building the tracks, the BFS algorithm further identifies the end-point voxels of each track. These are defined as the pair of voxels with the longest distance between them, where the distance between any pair of voxels is defined as the shortest path along the track that connects them. The SiPM hits contained in the end-point voxels are used to define the voxel center-of-gravity (COG). Finally, two spheres of a fixed radius are defined around the end-point voxels COGs, and the energy contained in them is summed, designating the sphere containing more energy as {\it blob1} and the one carrying less energy as {\it blob2}. The final output for analysis consists of a collection of tracks and their ``blobs''.

To investigate the effectiveness of topological event classification in NEXT-White, the experimental work described in \cite{Ferrario:2019kwg}, to which we refer below as the {\it classical analysis}, focused on $e^-e^+$ pairs produced in Xe by the 2615\;keV gamma of $^{208}$Tl, using an external source of $^{228}$Th (parent of the decay chain containing $^{208}$Tl). The topological analysis of such pairs provides an excellent ``training arena'' for that of $0\nu\beta\beta$ events. Similar to the track structure of $0\nu\beta\beta$ decay, the pair electron and positron tracks start from a common vertex, and both end with a Bragg peak of dense ionization. Such pairs have a total kinetic energy of 1593\;keV, and their selection is based on the identification of the $^{208}$Tl double escape peak in the energy spectrum. At 10\;bar the combined continuous-slowing-down approximation range of such pairs is $\sim20$\;cm, the same as that of $0\nu\beta\beta$ events at 15\;bar (the planned operation pressure of NEXT-100); the typical projected length of the full event in both cases is $\lesssim$10\;cm. For the sake of the analysis, $e^-e^+$ pairs are therefore considered as signal events, while single-electron events (from Compton scatters) in the region of the double escape peak, are considered as background.

For the classical analysis the SiPM charge cut was set at a high value of 30 PEs, SiPM hits were grouped in cubic voxels with an event-dependent fixed size between 10 and 15\;mm (figure 2, right), and the blobs were defined by 21\;mm-radius spheres. All choices are revisited in the present work. In the analysis (classical and the new method presented here), events are required to be fully contained within the fiducial volume (with all hits at least 2 cm away from any of the drift volume surfaces) and consist of only one track. Classification to signal or background is done by comparing the energy contained in blob2 to a fixed threshold. If blob2 energy exceeds it the event is classified as signal, and if not -- as background.

To quantify the effectiveness of this topological cut on double escape peak events, one defines the {\it signal efficiency} $\epsilon$ as the fraction of signal events ($e^-e^+$ pairs) passing the cut, and the {\it background acceptance} $b$ as the fraction of background events that survive it. Both parameters depend on the choice of the energy threshold for blob2. As discussed in \cite{NEXT100_sensitivity}, the sensitivity of the experiment to $0\nu\beta\beta$ decays is proportional to a topological figure of merit, defined as $f.o.m. = \epsilon/\sqrt b$, where the signal efficiency and background acceptance are calculated for events in the $Q_{\beta\beta}$ ROI. Considering the similarity between 1.6\;MeV $e^-e^+$ pairs at 10\;bar and $0\nu\beta\beta$ events at 15 bar, optimizing the $f.o.m.$ for $e^-e^+$ events -- which are accessible experimentally at large numbers -- can be regarded as a proxy for optimizing the sensitivity of the experiment to $0\nu\beta\beta$ decays. For a given configuration of reconstruction parameters, the maximal $f.o.m.$ is obtained for an optimal choice of blob2 energy threshold. As shown in \cite{Ferrario:2019kwg} for the optimal figure of merit, the classical analysis, when applied to experimental data, yields a signal efficiency of 71.6\% with a background acceptance of 20.6\% (with small statistical and systematic errors), i.e., a background rejection factor of 4.9. Similar outcomes were obtained in Monte Carlo (MC) simulations of double-escape peak events, performed with the NEXUS Geant4-based NEXT simulation framework \cite{Justo_thesis}. When evaluating the performance of the classical analysis at $Q_{\beta\beta}$ using MC events, the background acceptance was further reduced down to 13.6\% (background rejection factor of 7.4) while keeping virtually the same signal efficiency \cite{Ferrario:2019kwg}.

As mentioned earlier, a more recent work \cite{Kekic:2020cne} improved on these results by using a deep convolutional neural network to classify 1.6\;MeV double-escape peak events to signal and background, starting from the voxelized tracks, instead of using the BFS algorithm and blob-based analysis. This method yielded a background acceptance level of 10\% at 65\% signal efficiency. The reconstruction method described in the following sections is closer in nature to the classical analysis, which therefore serves as the reference for comparison. Future works will investigate the benefit of combining image deconvolution and neural network-based analysis.

\begin{figure}
	\begin{center}
		\includegraphics[trim={0 0 0 0}, clip, height=6cm]{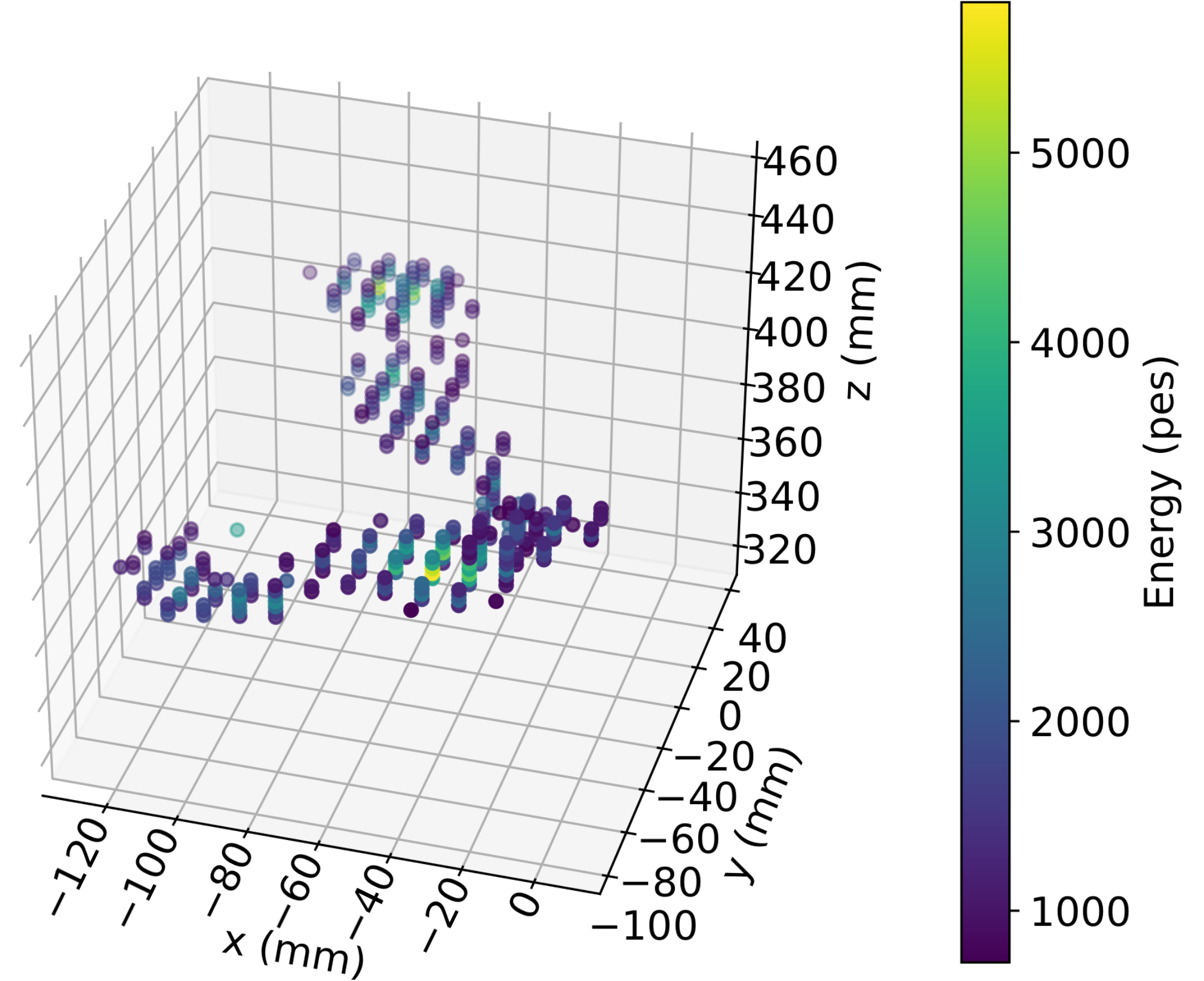}
		\includegraphics[trim={0 0 0 0}, clip, height=6cm]{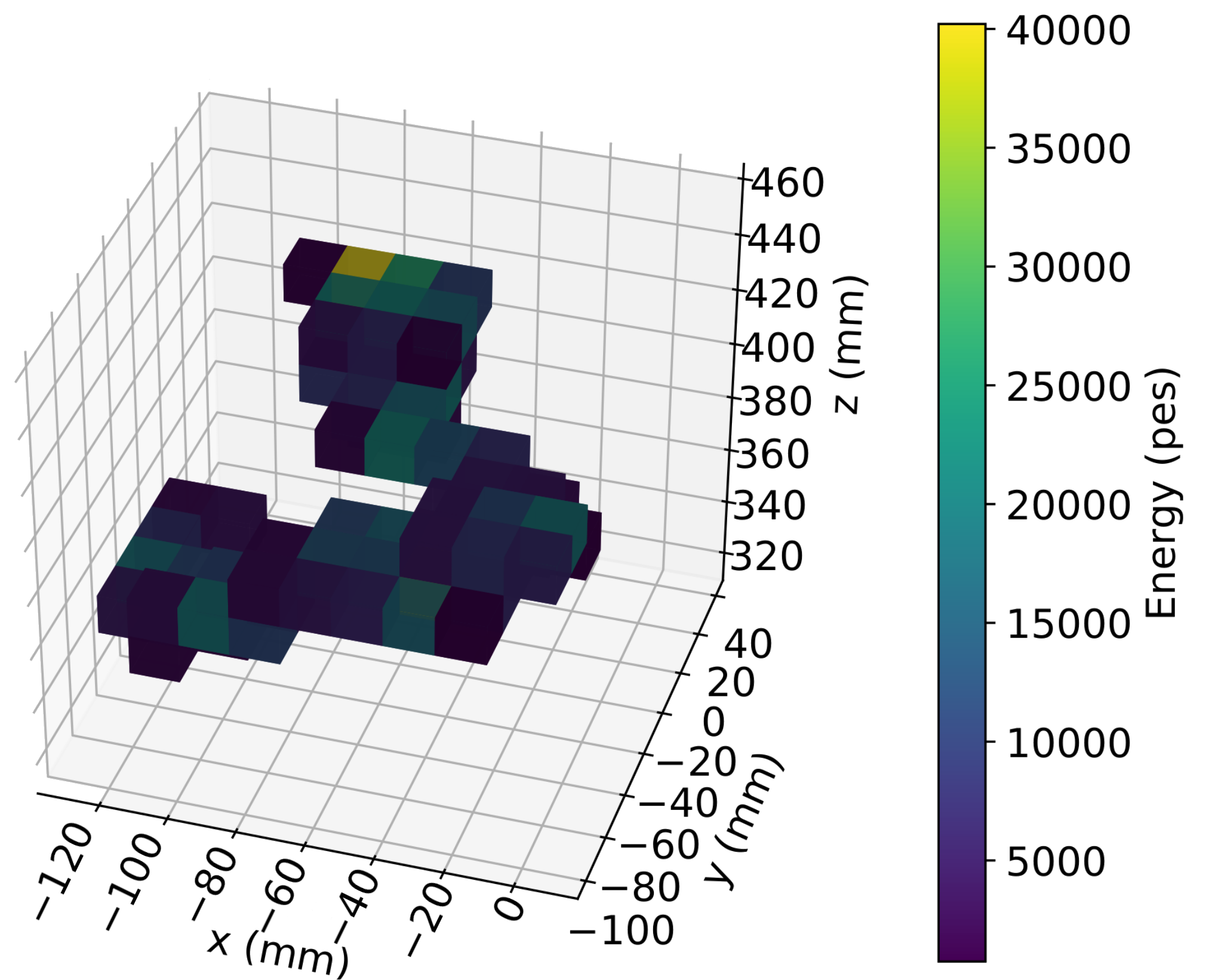}		
		\caption{An example for a reconstructed event in NEXT-White. Left: SiPM hits, with color representing detected photons; right: same event voxelized in $15\times15\times15$\;mm$^3$ bins (from \cite{Ferrario:2019kwg}). }
		\label{3D_hits}
	\end{center}
\end{figure}

%% file: src/blurring.tex
\section{Image blurring effects in NEXT-White}
\label{sec:blurring}

The classical reconstruction using the SiPM hits and voxels is affected by two blurring mechanisms which degrade the quality of the reconstructed track: electron diffusion and the spread of EL light on the tracking plane.

The initial track structure is a thin trail of ionization, as shown in figure \ref{img:geant4_tracks}. As the ionization electrons drift towards the gate, elastic collisions with Xe atoms lead to transverse and longitudinal diffusive spread of the charge cloud around the track ``backbone''. The root-mean-square (r.m.s.) diffusive spread of each point-like element of the initial track is proportional to the square root of the drift time, and therefore -- since the drift velocity is constant -- to the square root of the distance between this element and the gate (i.e., its $z$ coordinate). Under the operating conditions of NEXT-White, this effect can be on the cm scale.

The second contribution to the overall blurring occurs as the electrons cross the EL gap, where they emit light isotropically in $4\pi$. VUV photons emitted towards the anode are absorbed in TPB and re-emitted (again isotropically) as blue photons, which are subsequently transmitted through or reflected on the interfaces of the multi-layered anode plate, with additional reflections from the tracking plane PTFE cover. VUV EL photons emitted towards the gate and absorbed on it may result in secondary photoelectron emission, creating an additional discrete ``halo'' of diffuse light around the event. All of these processes combine to further optically smear the image of the charge distribution in the EL gap at a given time slice, with a similar relative contribution as electron diffusion. 

\begin{figure}
	\begin{center}
		\includegraphics[width=1.0\textwidth]{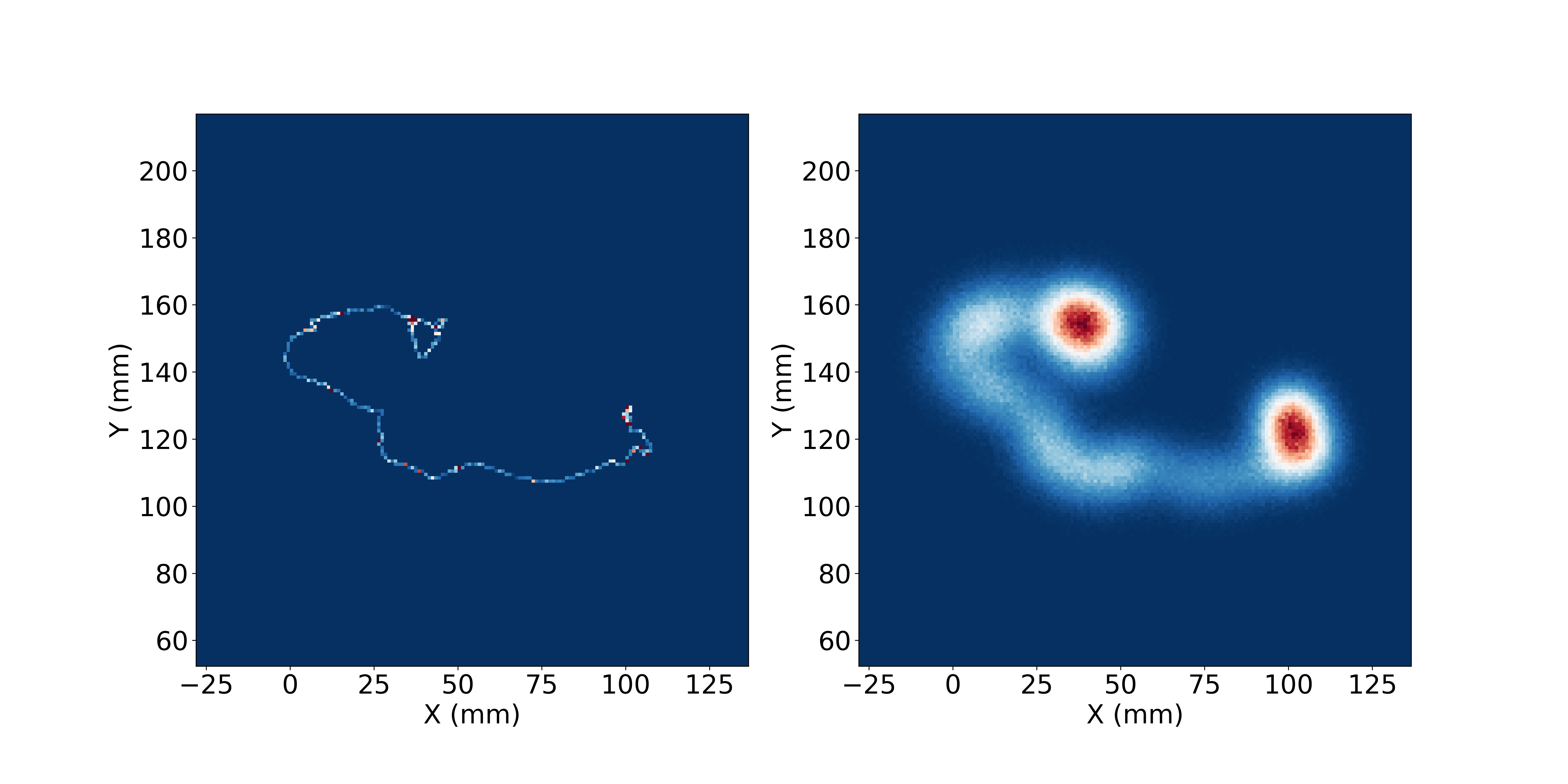}
		\caption{Simulated $^{136}$Xe $0\nu\beta\beta$ event. Left: initial track; Right - same track after drifting 40 cm considering the longitudinal and transverse diffusion coefficients to be 0.27\;mm$/\sqrt{\textrm{cm}}$ and 1.07\;mm$/\sqrt{\textrm{cm}}$ respectively. Color scale in the right plot represents the number of ionization electrons at each $xy$ bin.}
		\label{img:diffused_track}
	\end{center}
\end{figure}

Both blurring effects can be characterized using $^{83m}$Kr data. In fact, transverse and longitudinal diffusion in NEXT-White were already studied using $^{83m}$Kr events in \cite{Simon:2018vep}. The r.m.s. transverse diffusion spread at 10 bar was found to be 1.07\;mm${\times\sqrt{z(\textrm{cm})}}$ and the longitudinal one -- 0.27\;mm${\times\sqrt{z(\textrm{cm})}}$. For the full drift distance in NEXT-White, $z=53$\;cm, 
the transverse and longitudinal FWHM spread of an electron cloud starting from a point-like charge distribution are, in this case, 18.3\;mm and 4.6\;mm, respectively. An example for the effect of diffusion on a simulated $0\nu\beta\beta$ event for a drift distance of 40\;cm is shown in figure \ref{img:diffused_track}.

Both electron diffusion and the EL light spread can be quantified in terms of point spread functions. The full diffusion PSF is three dimensional: a point-like initial electron cloud transforms after diffusion to an oblate 3D Gaussian (wider in the transverse plane than along the drift direction), where both the transverse and longitudinal widths are proportional to $\sqrt{z}$. This 3D PSF can be projected on the $xy$ plane to yield an effective 2D transverse diffusion PSF, $F_{dif}^{2D}(x',y';z)$; (here $x'$ and $y'$ are the $xy$ coordinates in a frame of reference centered on the PSF axis). Similarly, integrating the total light hitting the tracking plane for a point-like charge crossing the EL gap produces a 2D EL PSF, $F_{EL}(x',y')$. Unlike the diffusion PSF, the EL PSF does not depend on the drift distance $z$. Detailed analysis of $^{83m}$Kr events show that except for the TPC edges, both the diffusion and EL PSFs do not depend, to leading order, on the absolute $xy$ position with respect to the TPC axis, and both are axisymmetric.

\begin{figure}[!h]
	\begin{center}
    \begin{minipage}{.5\textwidth}
    \centering
	\includegraphics[trim={20 50 30 55}, clip, height=6.3cm]{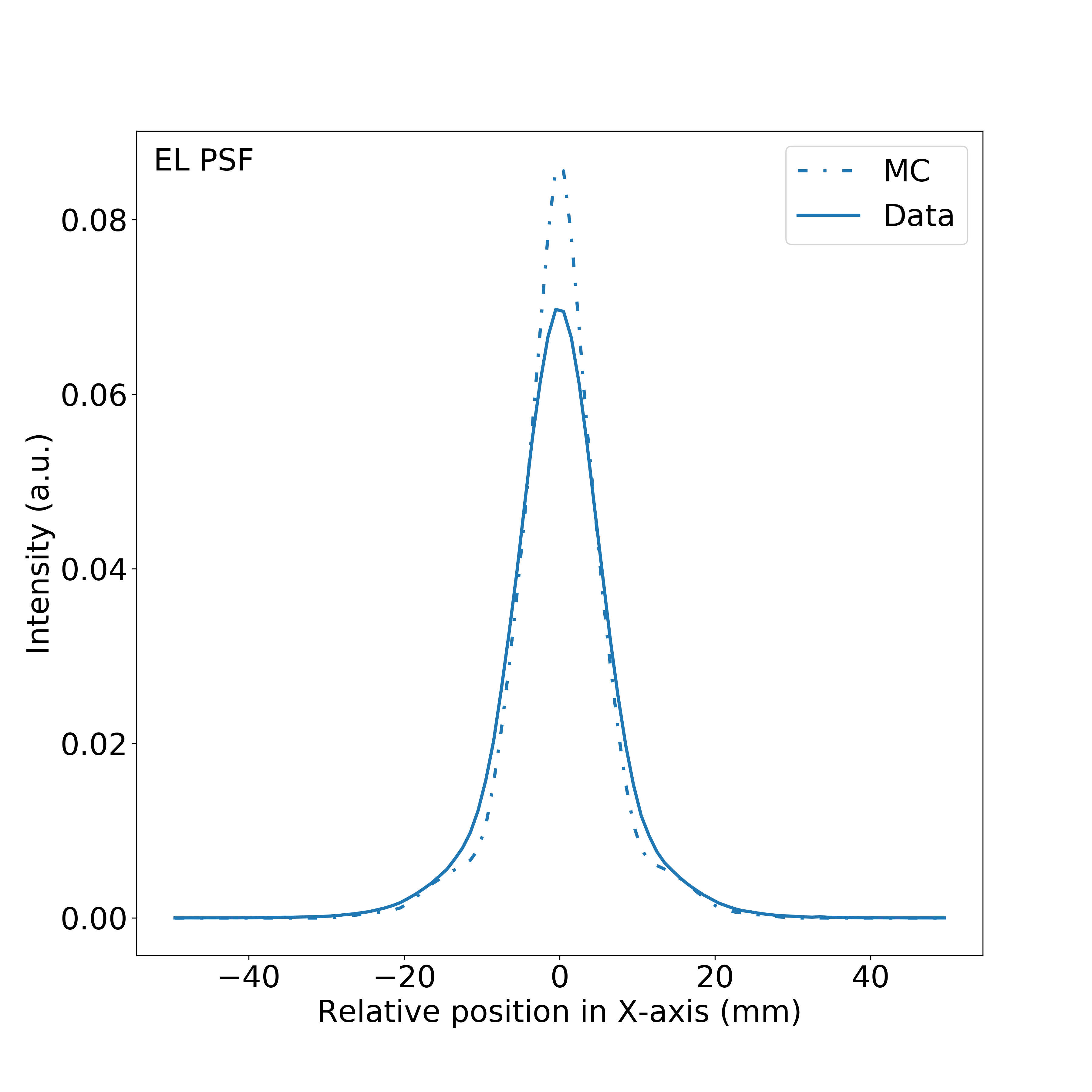}
    \end{minipage}%
    \begin{minipage}{.5\textwidth}
    \centering
	\includegraphics[trim={20 50 30 55}, clip, height=6.3cm]{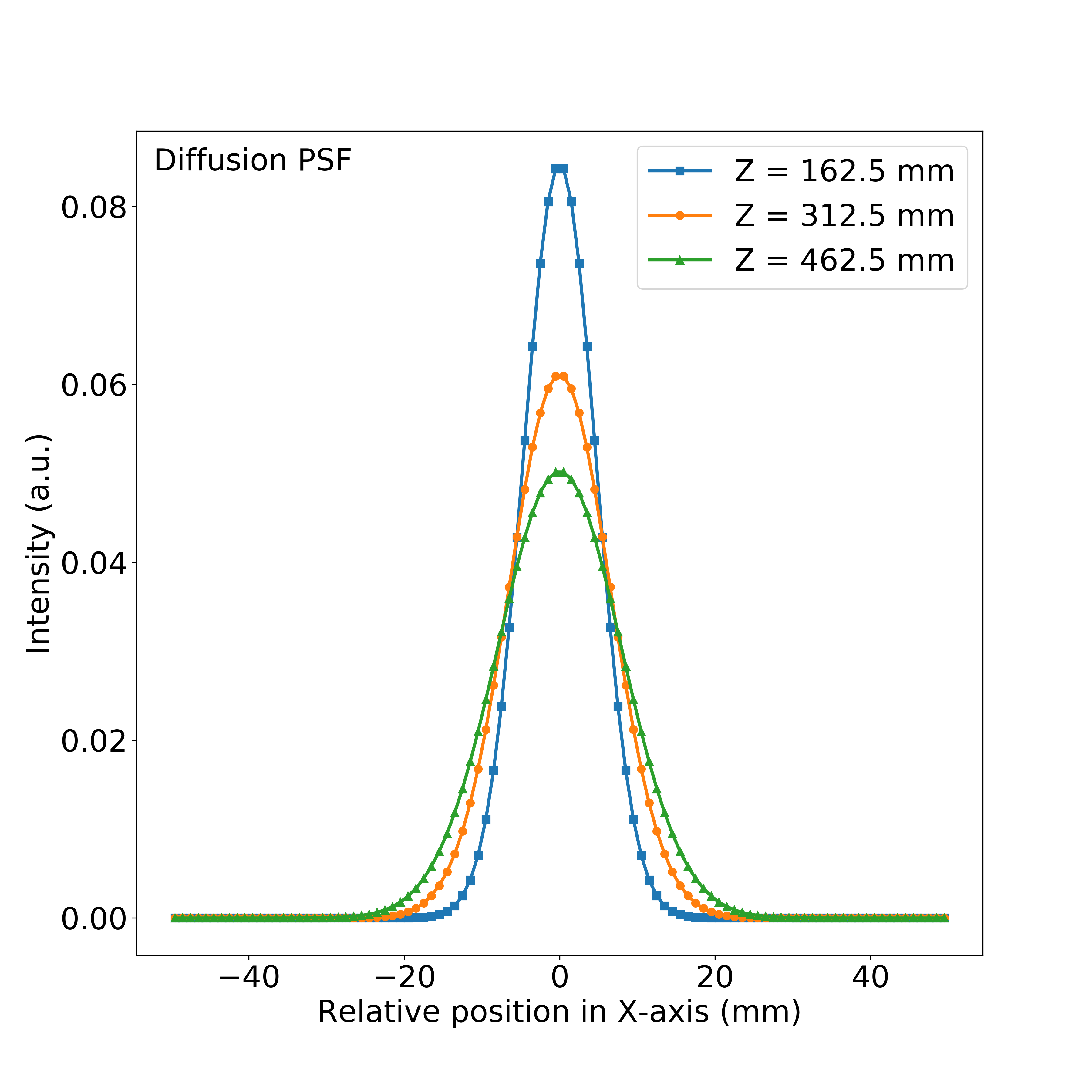}
    \end{minipage}%
    
    \begin{minipage}{.5\textwidth}
    \centering
	\includegraphics[trim={20 50 30 55}, clip, height=6.3cm]{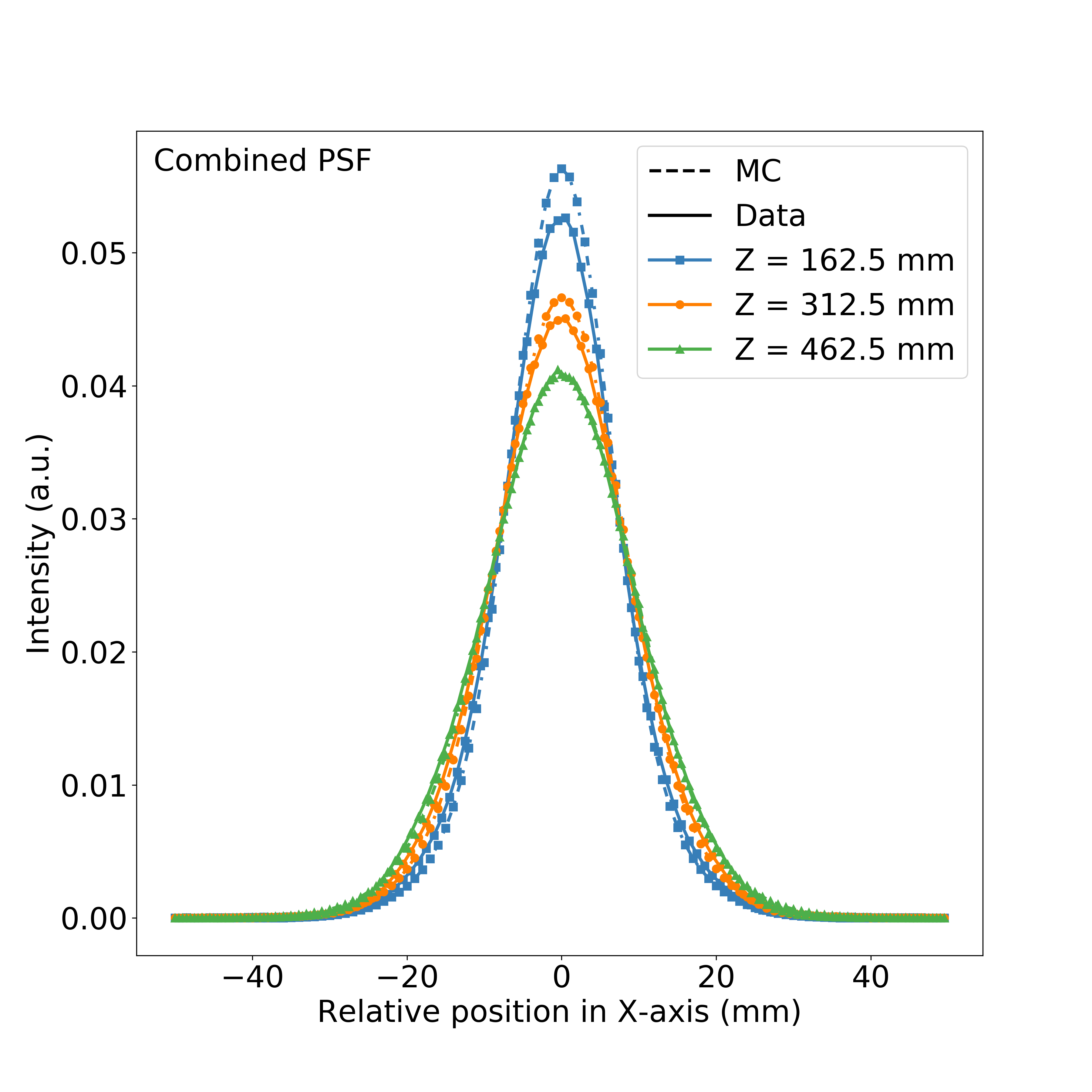}
    \end{minipage}%
    \begin{minipage}{.5\textwidth}
    \centering
	\includegraphics[trim={15 120 60 120}, clip, height=6.cm]{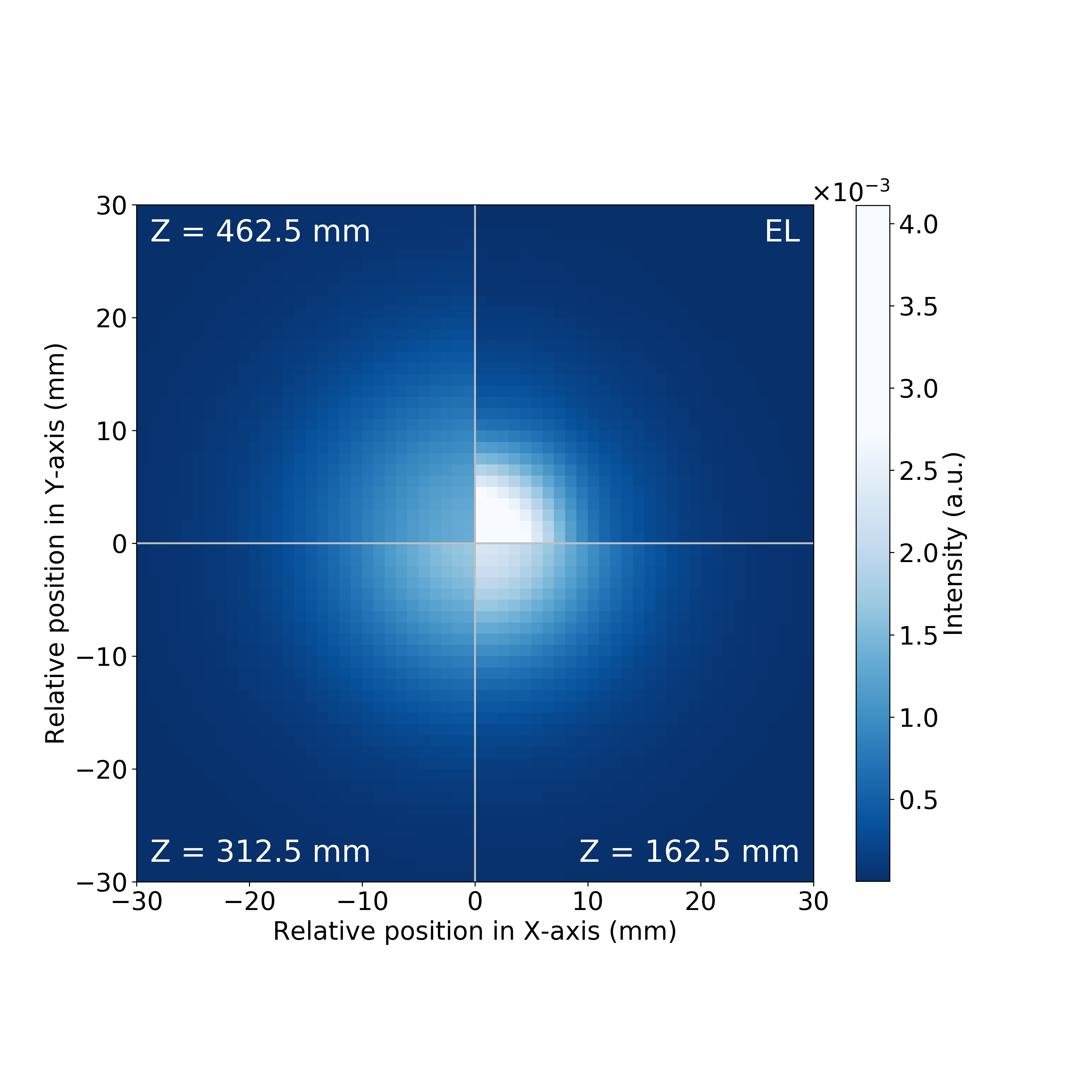}
    \end{minipage}
	\caption{Top left: EL PSF (from $^{83m}$Kr data and Monte Carlo). Top right: The 2D transverse diffusion PSF (calculation based on the diffusion measurements in \cite{Simon:2018vep}). Bottom left: The combined EL+diffusion PSF. Experimental data (solid line) are from $^{83m}$Kr events selected by their drift distance, while MC (dashed lines) is a convolution of the EL PSF and the 2D transverse diffusion PSF. Bottom right: The 2D PSF for experimental data for the different drifts shown in the accompanying plots, along with the 2D EL PSF in the first quadrant. The intensity value refers to the area-normalized value of the distributions.}
	\label{img:PSFs}
	\end{center}
\end{figure}

Experimentally, the EL PSF can be determined from $^{83m}$Kr events occurring immediately in front of the gate, such that they do not suffer a diffusive spread and can be considered point-like. The procedure, similar to that described in \cite{Simon:2018vep}, involves recording a large number of $^{83m}$Kr events over a small drift region (drift time $<25\;\mu$s). For each event, the SiPM response is integrated over several $\mu$s, to include the full S2 signal. The $xy$ location of the event is determined by calculating the center of gravity (COG) of the SiPM hit map. The coordinates of all SiPMs participating in the event are shifted to a reference frame whose origin coincides with the COG of the event, and their charge is binned in $1\times1$\;mm$^2$ pixels. The process, repeated over a large number of events, such that in each step the SiPM charge is added to the corresponding pixels, converges to the PSF. Figure \ref{img:PSFs} (top left) shows the profile of the EL PSF constructed from $^{83m}$Kr data. For comparison, we also show the PSF extracted from a Monte Carlo simulation. The MC PSF is narrower and the wings of the experimental PSF are somewhat larger. This indicates that the present simulation does not provide a complete description of the optical processes occurring in the EL gap and multi-layered anode plate (in particular, it does not include, at this stage, a contribution of secondary photoelectron emission from the gate, which may lead to single-electron EL signals at some distance from the main event).

The effective 2D transverse diffusion PSF is given approximately by a Gaussian whose standard deviation is: $\sigma_t=1.07\sqrt{z}$, where $z$ is in cm and $\sigma_t$ in mm. This function is shown in the top right panel of figure \ref{img:PSFs}, for several values of $z$. The two blurring effects of diffusion and EL light production can be combined in a single z-dependent PSF, which, to a good approximation, is given by their convolution: $F_{dif+EL}(z)=F_{dif}^{2D}(z) * F_{EL}$. This expression is only approximate, because in reality longitudinal diffusion introduces ``cross talk'' of charge between adjacent slices; however, since longitudinal diffusion is $\sim4$ times smaller than transverse diffusion, and is, in all cases, smaller than the width of the EL gap, we do not take into account this small effect. The combined PSF can be determined experimentally similarly to the EL PSF, by selecting $^{83m}$Kr events over a range $[z,z+\Delta z]$, as was done in \cite{Simon:2018vep} to determine the transverse and longitudinal diffusion coefficients. In the present study, events were selected in 25\;$\mu$s drift time intervals. This step size was chosen to keep variations in transverse diffusion within the interval sufficiently small. Thus, for the first 100\;$\mu$s drift, the relative change within a 25\;$\mu$s interval is $\sim$10\%, while at half the chamber, where the drift time is $\sim$225\;$\mu$s, the change falls below 5\%. The bottom panel of figure \ref{img:PSFs} shows the combined z-dependent PSF for the set of drift distances shown in the top right panel. We show both the experimental PSF and the one obtained by convolving the Monte Carlo EL PSF and the transverse diffusion Gaussian. As indicated by the figure, convolution of the EL PSF with the diffusion PSF washes out most of the differences between the two. Note that for the two different data sources, MC and detector acquired data, double escape peak events are analyzed below with their respective z-dependent PSFs.

%% file: src/RL3.tex
\section{Track reconstruction in NEXT-White using Richardson-Lucy deconvolution}
\label{sec:RL}

The Richardson-Lucy (RL) algorithm (also known as the Lucy-Richardson algorithm) was developed independently by W. H. Richardson \cite{Richardson:72} and L. B. Lucy \cite{Lucy:1974yx} in the early 1970's in the context of observational astronomy. Richardson focused his discussion solely on the recovery, by deconvolution, of an underlying sharp image from an observed blurred one. Lucy's work, in contrast, considered a more general case, where one seeks to recover an underlying frequency distribution from an observed one, with image restoration as a particular application. The method has subsequently become a main tool for image restoration in many scientific and engineering fields. The algorithm is iterative, generating a sequence of improved approximations for the underlying sharp image based on the observed blurred and noisy one, and the (presumably known) point spread function. It can be applied on an arbitrary number of spatial dimensions. Appendix \ref{sec:RL_app} outlines the procedure for two-dimensional images. Here we describe its implementation within the NEXT event reconstruction scheme and a series of evaluations performed to cross-check and validate the methodology. 

\subsection{RL implementation in NEXT-White}

In NEXT-White, we apply RL deconvolution on individual SiPM time-sliced hit maps (each integrated over a time interval $\delta t$, where, in the present analysis $\delta t=2$\;$\mu$s). Each slice is considered to be fully independent from the others and longitudinal spread is not taken into account. For a slice recorded at time $t$, we associate a physical slice of width $\delta z=v_d\delta t$ of the original 3D track at the corresponding drift distance $z=v_d\cdot(t-t_0)$ (for $v_d=0.9$\;mm/s, $\delta z=1.8$\;mm). We identify the physical slice, using the terminology of appendix \ref{sec:RL_app}, with the underlying sharp 2D image $W(x,y)$, and the corresponding SiPM hit map as a sampled representation of the blurred image $\tilde{H}(x,y)$. The two images are assumed to be related through the combined diffusion+EL PSF $F(x,y;z)$ corresponding to a drift distance $z$. 

The implementation of RL deconvolution on each slice is done, for both experimental data and MC events, in the following steps:

\begin{enumerate}
	
	\item Charge cut: SiPMs containing less charge than a predefined threshold $q_{cut}$ are removed from the slice. As discussed in appendix \ref{sec:RL_par}, after exploring several different choices for the charge threshold, we adopted a value $q_{cut}=10$\;PE (compared to 30 PE used in the classical analysis).
	\item Removal of isolated SiPMs: Single SiPM hits which have no adjacent non-zero neighbors in the same slice are removed. This is done to avoid filling the region between the main track and isolated SiPMs which fluctuate above the charge threshold $q_{cut}$ by non-physical data in the subsequent interpolation step.  
	\item 2D interpolation: We define a rectangular region surrounding the SiPMs which have survived steps (1) and (2) with 10\;mm margins. To estimate the full pattern of photon hit points in this region, we apply bicubic 2D interpolation on the ``cleaned up'' SiPM hit map over a $1\times 1$\;mm$^2$ grid (the SiPMs cover only 1$\%$ of the plane). Note that no significant differences were observed in the final outcomes of the analysis (signal efficiency and background acceptance for double escape peak events) when replacing bicubic by linear interpolation (see a brief discussion in appendix \ref{sec:RL_par}).
	\item RL deconvolution: For each slice, we use the corresponding z-dependent combined EL+diffusion PSF for the deconvolution process, following equations (\ref{eq: H(r)})-(\ref{eq: W(r)}) in appendix \ref{sec:RL_app}, to find successive estimations $W^{(r)}(x,y)$ for $W(x,y)$ in $N_{iter}$ iterations. Data and MC PSFs are used for data and MC events, respectively. The process maintains the overall charge of each slice constant in all iterations. It was implemented using the Richardson-Lucy function from Python's scikit-image library \cite{scikit-image}.
	\item Cleaning cut: Once the iterative process is completed, a cleaning cut with an adjustable threshold $\epsilon_{cut}$ is applied to the image intensity given by the iterative process. This is done to remove non-physical backgrounds and reconstruction leftovers, and sharpen the track edges for the topological analysis. For the double escape peak analysis the cut was set at 0.008 a.u. Details on the optimization can be found in appendix \ref{sec:RL_par}. No cut was performed when applying the method to Kr events as the reconstruction leftovers where not found to have an impact on the performance.
	\item Energy allocation: Finally, based on the integrated S2 signal recorded by the PMTs over the entire event duration, and using the $^{83m}$Kr-based lifetime and S2 correction maps \cite{Renner:2019pfe}, we find the total energy of each recorded slice and divide it among all of the $1\times 1$\;mm$^2$ pixels of the deconvolved image, proportionally to their interpolated charge.
\end{enumerate}

The first three steps aim to generate a {\it reasonable} estimate for the actual photon hit pattern on the tracking plane. They reflect a pragmatic approach to bridge the empty spaces (and hence lack of information) between the SiPMs, and to avoid distorting the image by distant effects, such as reflections from the various TPC surfaces, or distant EL light emission by photoelectrons ejected from the gate mesh. The interpolation step is justified as the smearing effects of both electron diffusion and EL light spread produce gradual changes in light intensity on the tracking plane. Rather than claim for absolute mathematical rigor in this approximation, we provide a series of simple demonstrations to support its practical value. These include the reconstruction of individual Kr events, adjacent pairs of Kr decays, and straight muon tracks.

\subsection{Validation tests}

\begin{figure}
	\begin{center}
		\includegraphics[width=1.0\textwidth]{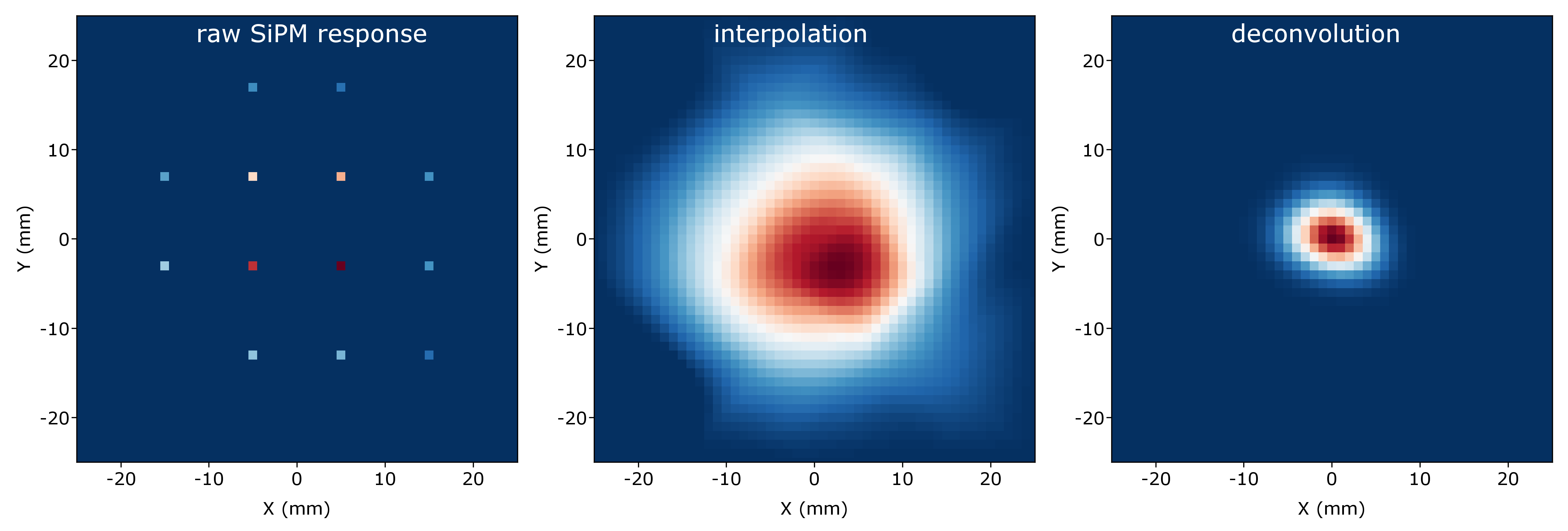}
		\caption{Example of a reconstructed $^{83m}$Kr event from NEXT-White data. The event is centered at (0,0) for convenience. Left: raw sensor response, with a charge cut of 10\;PE. Center: bicubic interpolation. Right: deconvolved image after 75 RL iterations.}
		\label{img:Kr1}
	\end{center}
\end{figure}

As a first test, RL deconvolution was applied to individual Kr events from both MC and detector data. A typical example, from data, is shown in figure \ref{img:Kr1}, with the SiPM sensor response for $q_{cut}=10$\;PE on the first column, bicubic interpolation on the second, and deconvolved images after 75 RL iterations on the third. When comparing the deconvolved images to the MC true information, the r.m.s. error in the reconstructed COG of all Kr-events was $\sim0.5-1.0$\;mm in both $x$ and $y$. 
For 75 RL iterations, the FWHM of the reconstructed Gaussian-like charge distribution was $\sim5$\;mm for both MC and detector data, independent of the drift distance, for $z>100$\;mm. 
Although adding iterations was found to reduce the FWHM further, the effect was quite marginal (e.g., applying 150 iterations reduced the FWHM to $\sim4$\;mm). 

\begin{figure}
	\begin{center}
		\includegraphics[width=1.0\textwidth]{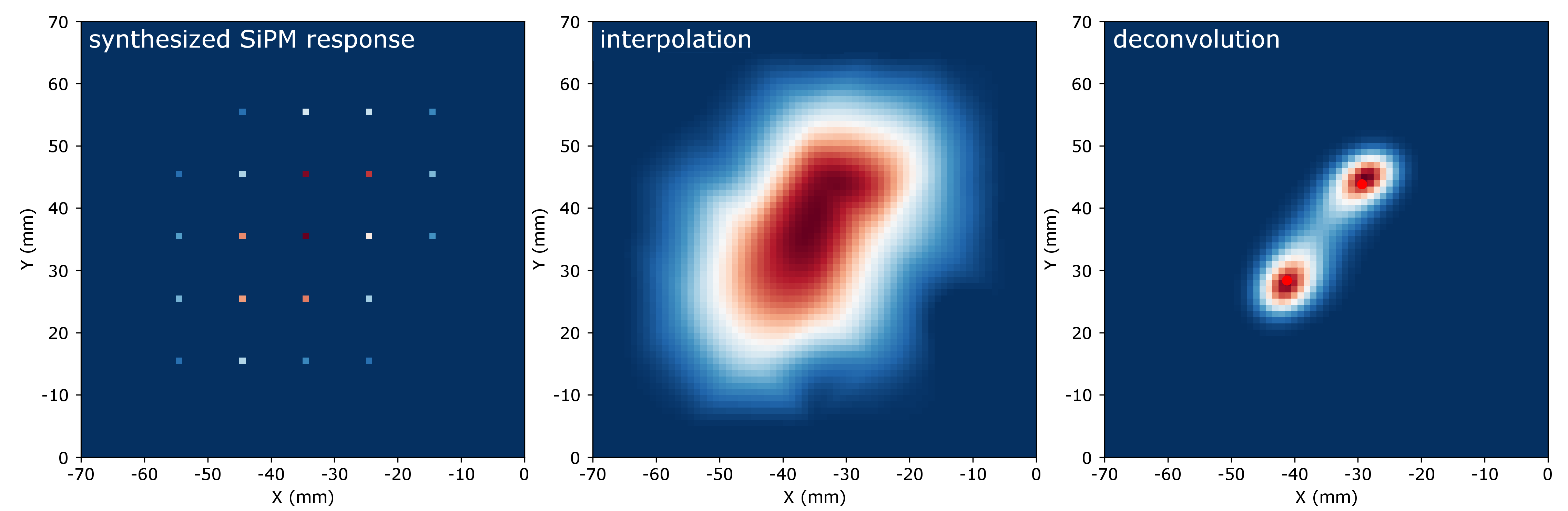}
		\caption{Reconstructed pair of nearby $^{83m}$Kr events from detector data, synthesized by overlaying SiPM response maps for two nearby events. Left column - raw (synthesized) SiPM data; center - bicubic interpolation; right - deconvolved image after 75 iterations. Red dots represent the COGs of individual events.}.
		\label{img:Kr pairs}
	\end{center}
\end{figure}

While the accuracy in COG reconstruction using RL deconvolution for point-like Kr events was practically the same as obtained from the raw SiPM response, the strength of the method lies in the ability to resolve nearby structures. To demonstrate this, we artificially synthesized pairs of Kr events from detector data, by overlaying SiPM sensor response maps for two individual events with the same drift distance. Image synthesis was done by shifting the SiPM response map of one event by an integer number of 10\;mm steps in $x$ and $y$, to bring its COG close to that of the other one. An example is shown in figure \ref{img:Kr pairs}, where the COGs are 19.5\;mm apart (for a drift distance of $\sim360$\;mm). The dots represent the individual Kr COGs. As in figure \ref{img:Kr1}, the left, center and right columns show the raw (overlaid) sensor response maps, interpolated images and deconvolved ones (with 75 iterations).

\begin{figure}
	\begin{center}
		\includegraphics[width=1.0\textwidth]{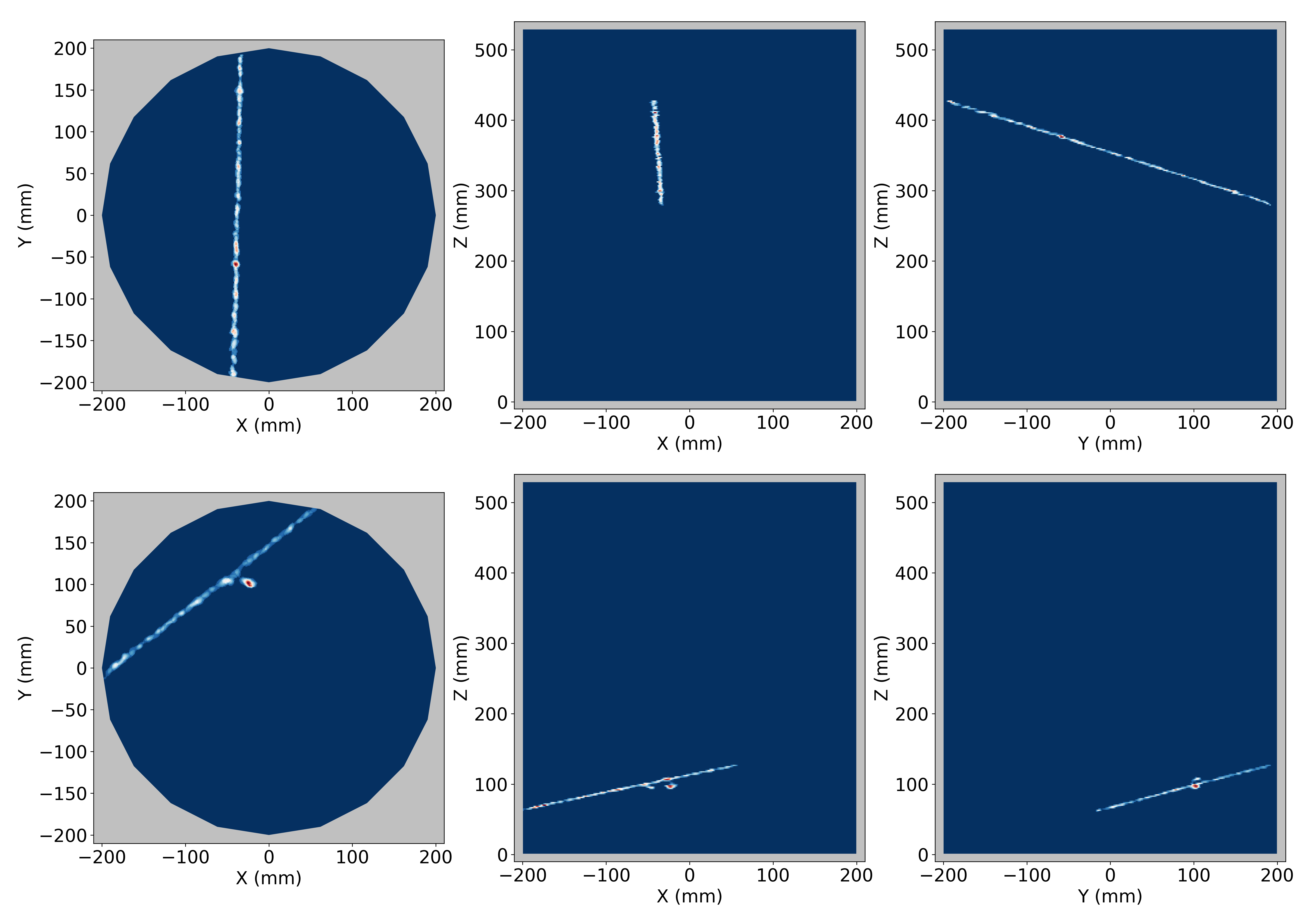}
		\caption{Two muon events in NEXT-White data after RL deconvolution, shown in three Cartesian projections. The top event is of a ``clean'' muon, while the bottom one also contains a delta electron.}
		\label{img:muon_tracks}
	\end{center}
\end{figure}

To demonstrate the method over long tracks, we selected, from detector data, a set of crossing muon events. A visual inspection was carried over an extensive dataset with no observable deviations from the expected straight line tracks, with occasional delta electrons branching out from the main track. Two examples of muon tracks of NEXT-White data (Run V) are shown in figure\;\ref{img:muon_tracks}.

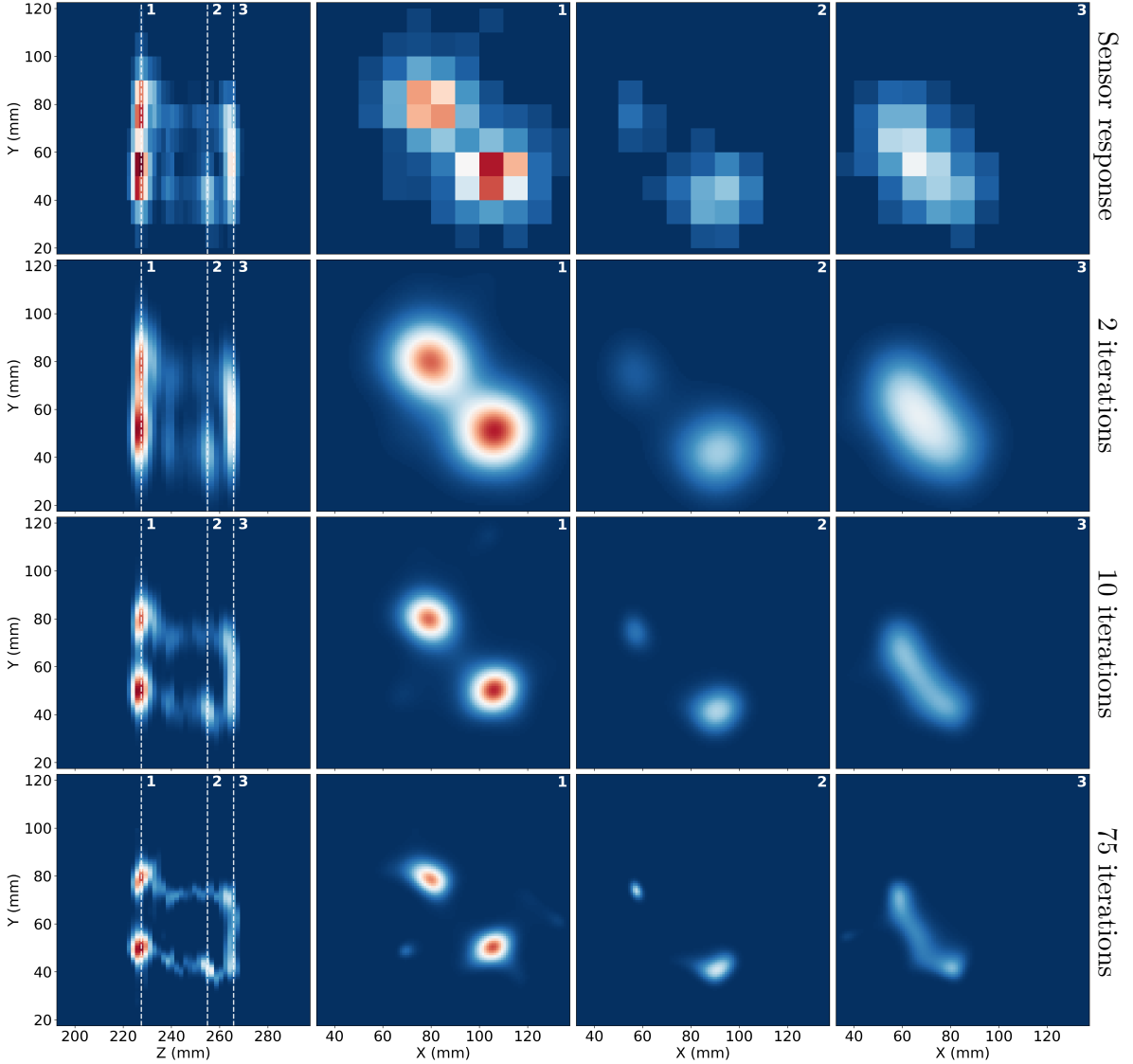
\begin{figure}
	\begin{center}
		\includegraphics[trim={7.4cm 8.3cm 16cm 16.2cm}, clip, width=1.0\textwidth]{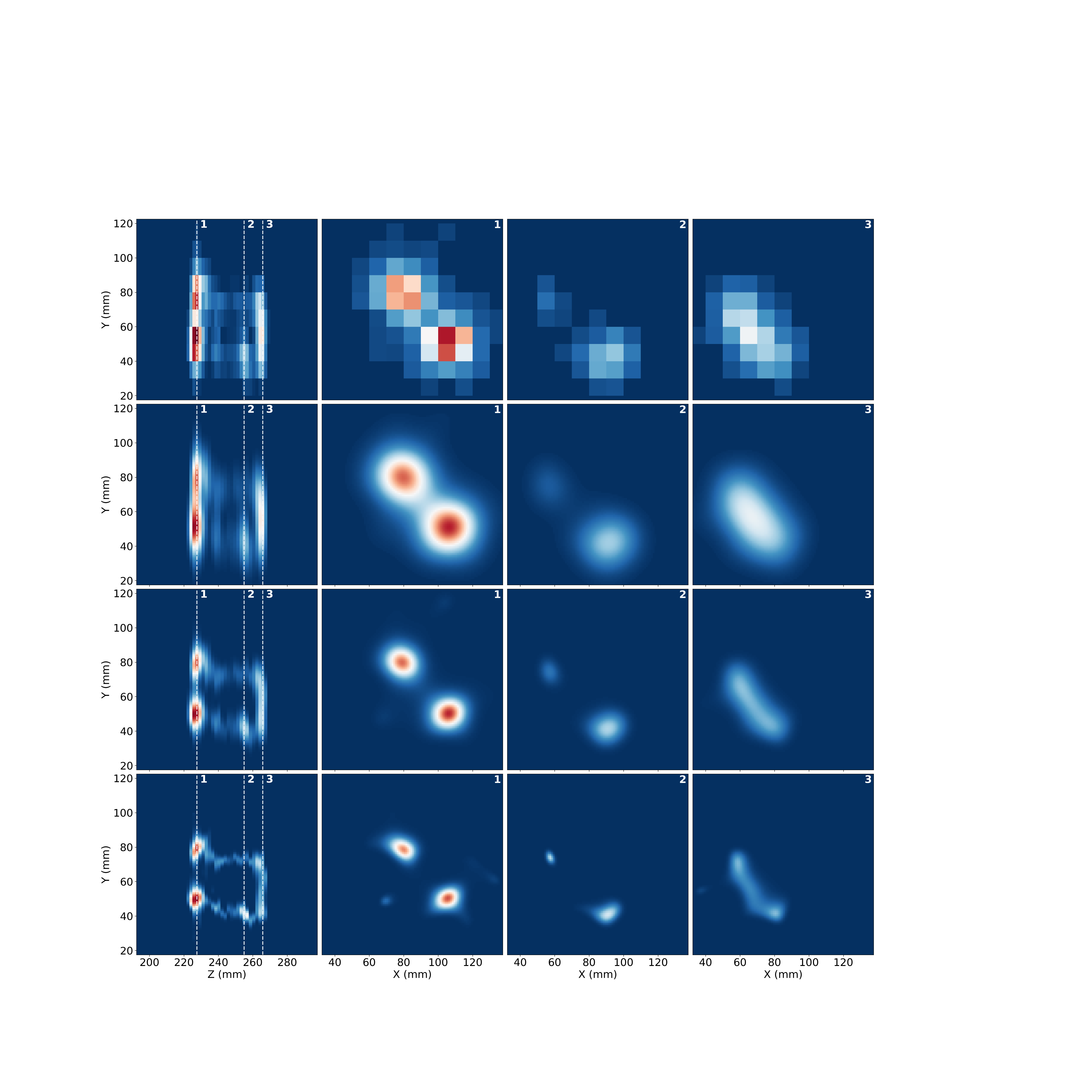}
		\put(0,410){\rotatebox{270}{Sensor response}}
		\put(2,296){\rotatebox{270}{2 iterations}}
		\put(2,196){\rotatebox{270}{10 iterations}}
		\put(2,93){\rotatebox{270}{75 iterations}}
		\caption{Effect of successive RL iterations (2, 10 and 75) in three selected slices. The event is an electron-positron pair (from data).}
		\label{img:RL_iterations_slices}
	\end{center}
\end{figure}

Figure \ref{img:RL_iterations_slices} demonstrates the iterative refinement in the track sharpness obtained by the RL process. The event consists of a 1.6\;MeV electron-positron pair, acquired in NEXT-White, which forms a U-shaped track, where the two ends are located roughly in the same $xy$ plane. The first row shows the raw data (sensor response) before deconvolution, binned in $10\times 10\times 1.8$\;mm voxels. The full track, projected on the $yz$ plane, is shown on the top left. The dashed lines mark three particular slices where the effect of applying successively more iterations is shown for each one individually. The corresponding slice images are shown in columns 2-4. Rows 2-4 show the same event after 2, 10 and 75 iterations, with the latter after application of the cleaning cut. 
The disconnected artifact appearing in slice 1 after 75 iterations (around $x=70$\;mm, $y=50$\;mm) does not affect the analysis of the main track, as explained in the following section.

%% file: src/pp_analysis.tex
\section{Analysis of 1.6 MeV e$^-$e$^+$ events}
\label{sec:pp_analysis}

\subsection{Methodology}

The detector data used for this analysis were taken during August 2019 under the same conditions as in \cite{Ferrario:2019kwg}, but with a much longer electron lifetime of $\sim9$\;ms. The MC simulation dataset was exactly the same. Detector data consisted of events generated by a $^{228}$Th source placed in an external calibration port above the center of the drift region, and of internal $^{83m}$Kr events.


Similarly to \cite{Ferrario:2019kwg}, event selection for the topological analysis around the $^{208}$Tl double escape peak was done by applying the following filters: (1) event energy in the range $1.4-1.8$\;MeV; (2) full containment of the event in the fiducial volume, with all hits at least 2\;cm away from all borders of the drift region; (3) the event comprises only a single track, when binned into 15\;mm voxels. The latter filter, to which we refer below as the ``gross single-track cut'' (given the large voxels size), was employed on the SiPM data before deconvolution rather than at the end of the process, to avoid artificial track multiplicity which may be created during the RL iterations and as a result of the final cleaning cut. 

Events passing the above filters were processed according to the RL steps listed in section\;\ref{sec:RL}. The set of deconvolved 2D images (which, in fact, consisted of $1\times1\times1.8$\;mm$^3$ voxels with 1.8 mm along $z$) were re-binned into larger voxels of adjustable size. As discussed in appendix \ref{sec:RL_par}, 5-mm voxels provided the optimal results. The BFS algorithm was then used -- as in the classical analysis -- to combine the event voxels into one or more tracks. At this stage, events may have contained more than one track either because they consisted of multiple nearby physical tracks which were unresolved by the gross single-track cut, or due to artificial breaking of a single track into smaller segments during the RL process. Concretely, it was estimated, based on the true information of simulated data, that 29\% of the events passing the gross single-track cut were actually multi-track events. At the same time, for the optimal parameter configuration described below, 27\% of the deconvolved simulated events exhibited more reconstructed tracks than the number of segments in the true information. These additional ``satellites'' were of low intensity, with essentially no effect on the blob-based analysis.

Once the voxels were grouped into tracks, the BFS algorithm was used to find the track ends. Those of the longest track of the event served as the centers for two spherical blobs of a variable radius. 
As before, we defined blob1 as the more energetic blob and blob2 as the less energetic one. We set a threshold $E_{0,blob2}$ on blob2 energy such that events above it were considered signal and below it -- background. For the $i$-th value of blob2 energy threshold $E_{0,blob2}^{(i)}$ we defined the signal efficiency $\epsilon_i$, background acceptance $b_i$ and figure of merit $f.o.m_i$ as:

\begin{equation}
\epsilon_i=\frac{\mathrm{number\;of\;signal\;events\;with\;E_{blob2}>E_{0,blob2}^{(i)}}}{\mathrm{total\;number\;of\;signal\;events\;with\;no\;cut}} \label{eq:eps true}
\end{equation} 

\begin{equation}
b_i=\frac{\mathrm{number\;of\;background\;events\;with\;E_{blob2}>E_{0,blob2}^{(i)}}}{\mathrm{total\;number\;of\;background\;events\;with\;no\;cut}} \label{eq:b true}
\end{equation} 

\begin{equation}
f.o.m_i=\frac{\epsilon_i}{\sqrt{b_i}} \label{eq:fom true}
\end{equation}

For MC, the true nature of the event is known: any event containing a positron is signal, and any event without one is background. For each threshold $E_{0,blob2}^{(i)}$, the signal efficiency, background acceptance and figure of merit are found directly from equations (\ref{eq:eps true})-(\ref{eq:fom true}) using the true information. For experimental detector data, however, the nature of the event is unknown {\it a priori} and one must resort to a different approach, which involves fitting the ROI data around the double escape peak with an expression that describes both signal and background, before and after the application of the cut $E_{blob2}>E_{0,blob2}^{(i)}$ \cite{Ferrario:2019kwg}.

\begin{figure}
	\centering
	\includegraphics[trim={3.5cm 0.5cm 4cm 2cm}, clip, width=1.0\textwidth]{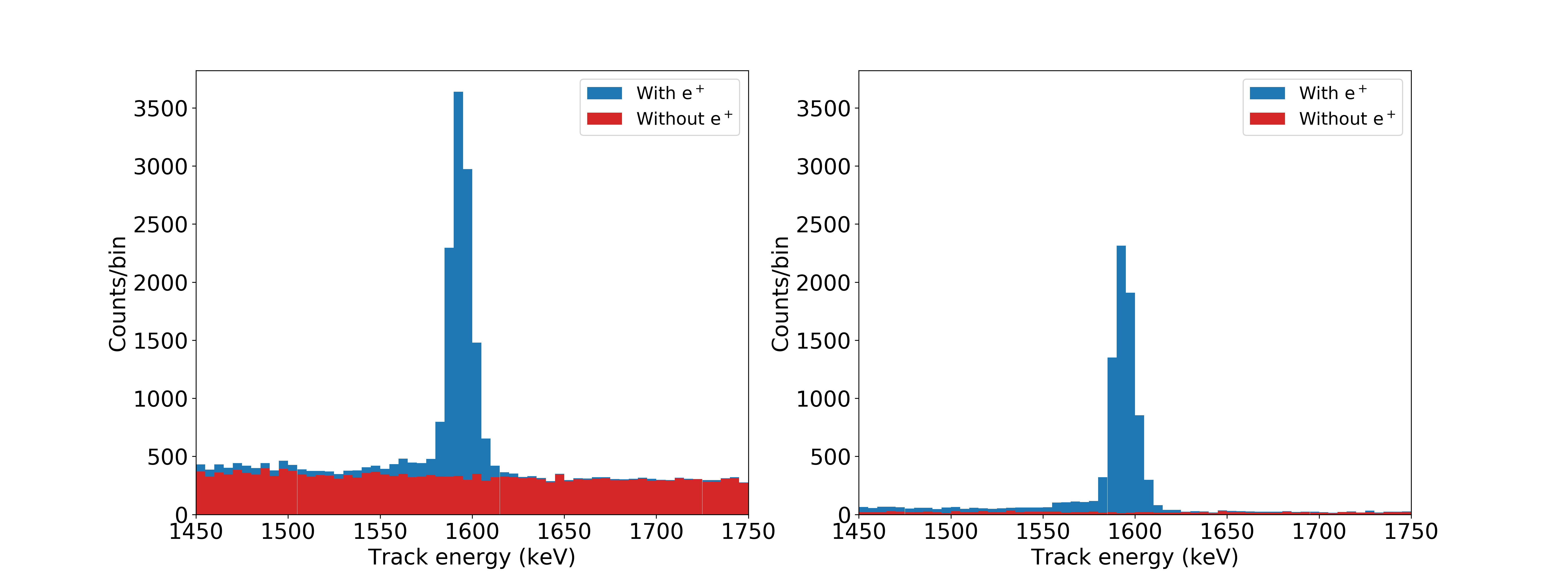}
	\caption{Energy spectrum and event population in the double escape peak region for MC events. In blue, events with a positron, in red -- the rest of the interactions (predominantly Compton electrons). Left: the spectrum before applying any topological cut. Right: after a blob2 energy cut of 340 keV.}
	\label{img:spectrum}
\end{figure}

Figure \ref{img:spectrum} shows the MC energy spectrum around the double escape peak, highlighting separately the contribution of $e^-e^+$ pairs (in blue) and of events with no positron (in red). The figure shows the spectrum before and after the application of the topological cut. The background spectrum in this region can be well described by a decreasing exponential, as was done in \cite{Ferrario:2019kwg}:

\begin{equation}
f_{bkg}(E)=A_1 \exp(-A_2 E) \label{eq:f_bkg}
\end{equation}

\begin{figure}	\begin{center}
		\includegraphics[width=0.6\textwidth]{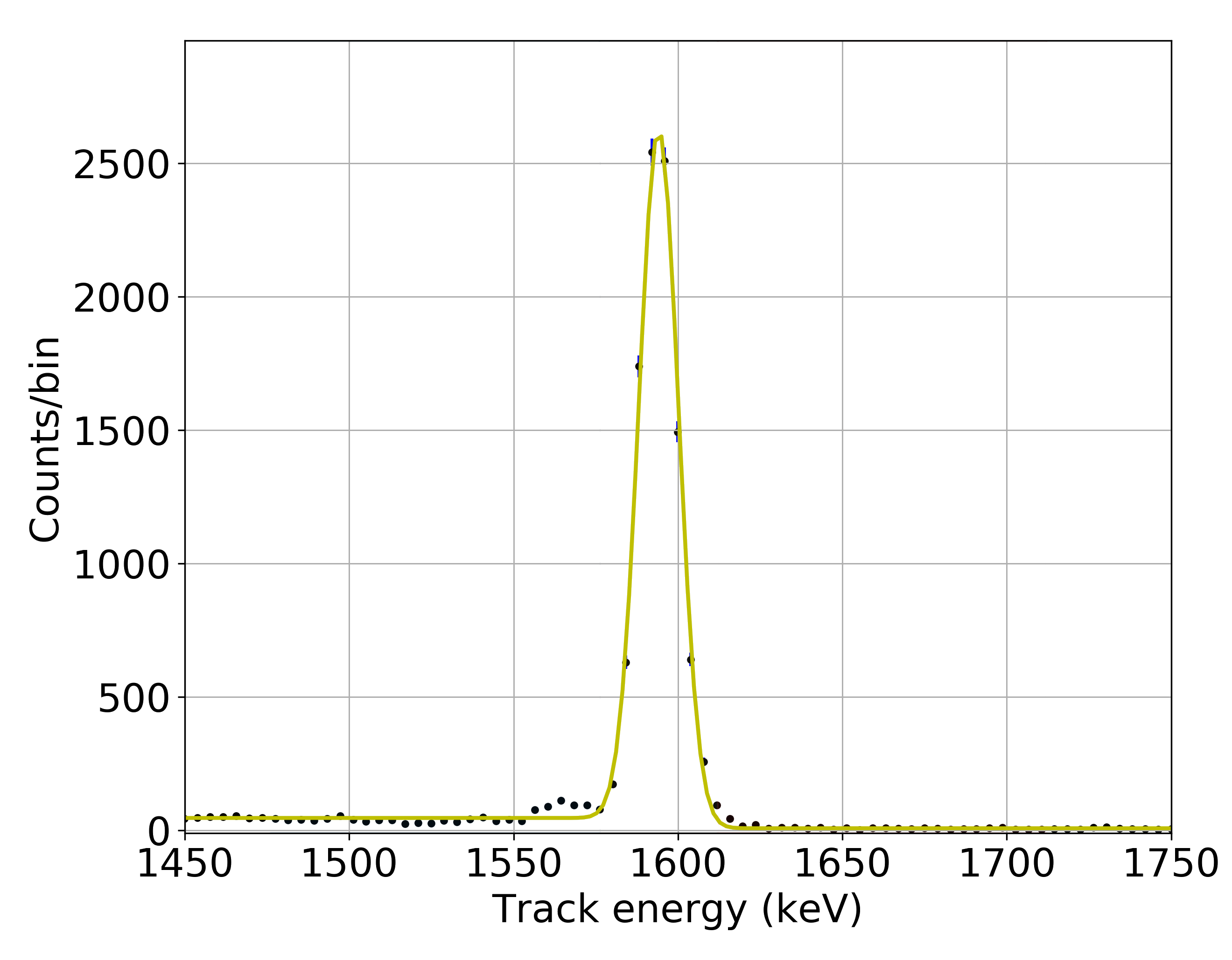}
		\caption{Energy spectrum of $e^-e^+$ pairs around the 1.6 MeV double escape peak, fitted according to equation (\ref{eq:f_sig}). }
		\label{img:signal_true_spectrum}
	\end{center}
\end{figure}

The signal spectrum consists of a Gaussian centered at 1593\;keV, with flat shallow wings for energies below and above the peak, as shown in figure \ref{img:signal_true_spectrum}. Below the peak, signal events comprise $e^-e^+$ pairs created by photons of energy below 2615\;keV (which predominantly result from Compton scatters of 2615\;keV gammas prior to pair production), as well as $e^-e^+$ pairs which lose some energy by bremsstrahlung, where the emitted photon does not interact in the sensitive volume. Above the peak there is a smaller population of $e^-e^+$ events where one of the 511\;keV gammas created when the positron annihilates interacts close to the main track and is unresolved by the gross single-track cut. The signal spectrum can therefore be approximated as:

\begin{equation}
f_{sig}(E)=B_1\left( \frac{1}{\sqrt{2\pi}\sigma}\exp\left( -\frac{\left( E-\mu\right)^2}{2\sigma^2} \right) + C_1\mathrm{erfc}\left(\frac{E-\mu}{\sqrt{2}\sigma}\right) + C_2  \right) \label{eq:f_sig}
\end{equation}
where a complementary error function (erfc) with the same standard deviation as the Gaussian is used to describe events below the peak. This expression was chosen empirically based on the MC distribution shown in figure \ref{img:signal_true_spectrum} and should be considered only as a proxy to obtain an estimate of the signal population outside the peak with no other physical meaning. The parameters $C_1$ and $C_2$ were extracted from fitting, through an unbinned extended maximum likelihood fit, MC data consisting only of $e^-e^+$ pairs in the region $1.4-1.8$\;MeV, giving: $C_1 = (4.89 \pm 0.46) \cdot 10^{-4}$ and $C_2 = (1.99 \pm 0.28) \cdot 10^{-4}$. Note that the analysis in \cite{Ferrario:2019kwg} disregarded the shallow wings in the $e^-e^+$ spectrum, and instead assumed that the signal is completely described by a Gaussian. This assumption is valid when the signal population outside the peak is negligible compared to the background, which is the case before application of the topological cut (figure \ref{img:spectrum}, left). For a modestly effective topological analysis, this holds also after applying the cut. However, this consideration loses validity as the topological cut becomes more effective, and a realistic fit should consist of the sum of equations (\ref{eq:f_bkg}) and (\ref{eq:f_sig}):

\begin{equation}
	f(E)=f_{bkg}(E)+f_{sig}(E) \label{eq:f=f_bkg+f_sig}
\end{equation}

The procedure for estimating the signal efficiency and background acceptance using the fit was as follows. The parameters $C_1$ and $C_2$ were extracted from the MC fit to the true signal spectrum and assumed to hold also for detector data, without being affected by the application of the topological cut. This approximation, which reflects an assumption that the signal efficiency of the the blob cut is energy-independent over the range $1.4-1.8$\;MeV, was confirmed to hold on MC data. 
For each value of $E_{0,blob2}^{(i)}$ we then fitted the data with $f(E)$ (equation (\ref{eq:f=f_bkg+f_sig})), with $A_1$, $A_2$ and $B_1$ as free parameters, before and after applying the cut $E_{blob2}>E_{0,blob2}^{(i)}$ using an unbinned maximum likelihood fit (the parameters $\mu$ and $\sigma$ were extracted once from a fit to the Gaussian alone and kept constant in subsequent fits). The number of background and signal events passing a given value of $E_{0,blob2}^{(i)}$ were calculated by integrals over $\pm3\sigma$ around the peak centroid:

\begin{equation}
N_{bkg,i}=\int_{\mu-3\sigma}^{\mu+3\sigma} A_{1,i} \exp( -A_{2,i}E ) dE \label{eq: N_bkg,i}
\end{equation} 

\begin{equation}
N_{sig,i}=\int_{\mu-3\sigma}^{\mu+3\sigma} B_{1,i}\left( \frac{1}{\sqrt{2\pi}\sigma}\exp\left( -\frac{\left( E-\mu\right)^2}{2\sigma^2} \right) + C_1\mathrm{erfc}\left(\frac{E-\mu}{\sqrt{2}\sigma}\right) + C_2  \right) dE \label{eq: N_sig,i}
\end{equation} 

The estimate was found to be within 1\% of the real number of events for both signal and background populations in MC. 
The signal efficiency and background acceptance for the cut $E_{blob2}>E_{0,blob2}^{(i)}$ were calculated as:

\begin{equation}
\epsilon_i = N_{sig,i}/N_{sig,0} \label{eq: sig eff fit}
\end{equation}

\begin{equation}
b_i = N_{bkg,i}/N_{bkg,0} \label{eq: bkg acceptance fit}
\end{equation}
where the subscript ``0'' refers to no cut (i.e., $E_{0,blob2}=0$). The figure of merit for the $i$-th threshold was calculated, as before, by $f.o.m_i=\epsilon_i/\sqrt{b_i}$.

\begin{figure}[!h]
	\begin{center}
		\includegraphics[width=1.0\textwidth]{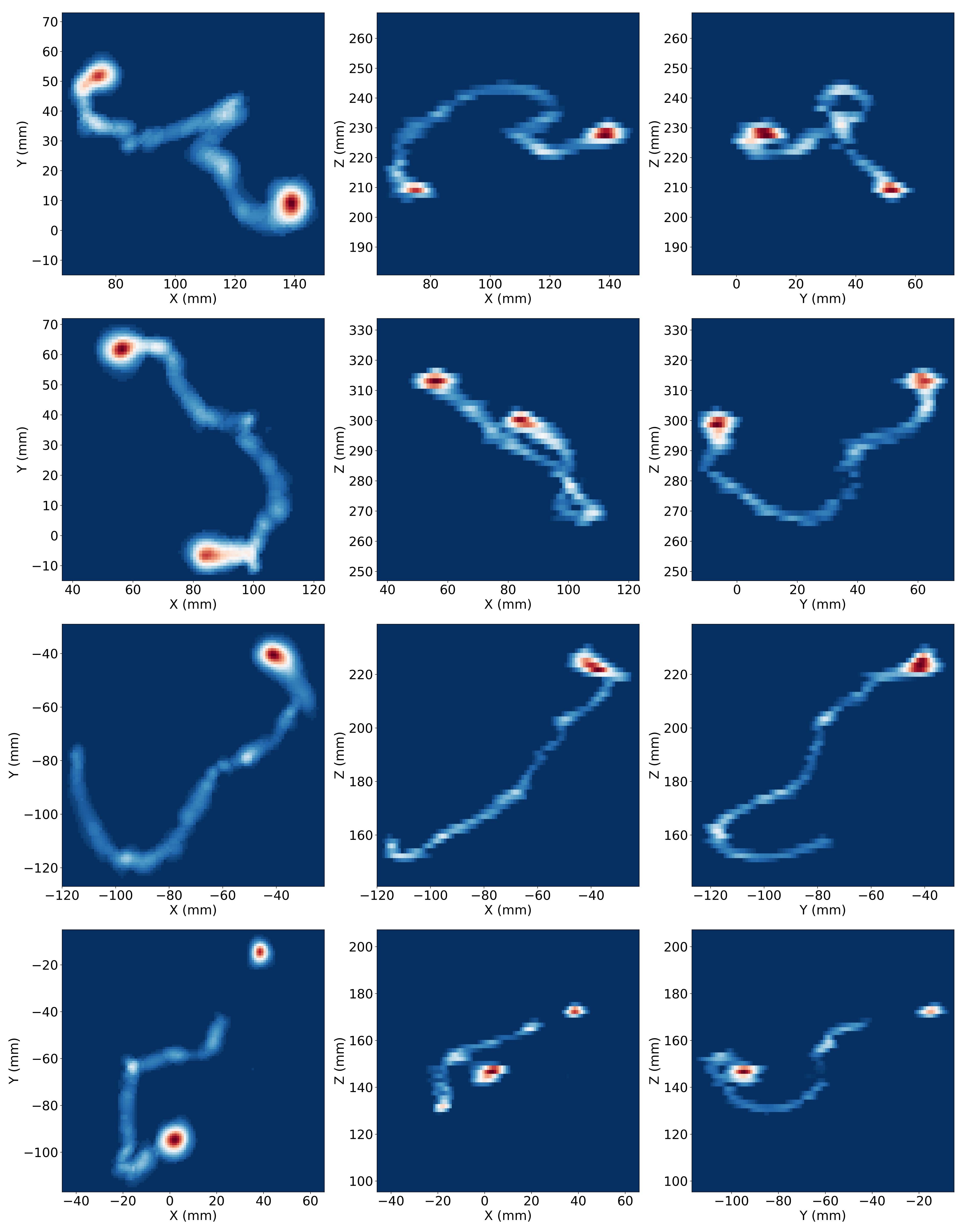}
		\caption{Experimental detector data events in the 1.6 MeV double escape peak region after topological classification. Top two rows: $e^-e^+$ candidates; bottom two rows: background candidates.}
		\label{img:double escape peak examples}
	\end{center}
\end{figure}

Figure \ref{img:double escape peak examples} shows four examples of detector data events reconstructed using RL deconvolution. The first two rows show events classified as signal, while the third and fourth rows display events classified as background (deconvolution and classification were done using the choice of parameters described below). Note that the second background event contains a blob-like satellite close to its start point. This event passed the gross single-track cut, and, in the classical analysis, would have likely been classified as signal. The refined reconstruction offered by RL deconvolution allows separating this satellite from the main track and correctly classify this event as background.

\subsection{Results} \label{sec:results}

The RL process and subsequent analysis involve several parameters, whose values determine the blob identification quality which ultimately reflects on the signal efficiency and background acceptance. These parameters can be divided in two groups: those affecting the quality of the reconstructed image and the end-point identification accuracy, and those which affect the energy calculation inside the blobs. The first group is related to the RL process and includes the SiPM charge threshold $q_{cut}$, number of RL iterations $N_{iter}$ and final cleaning threshold $\epsilon_{cut}$. The second group is comprised of the BFS characteristic parameters, namely the voxel size $l_{voxel}$ and blob radius $R_{blob}$. For a given choice of parameters the $f.o.m.$ attains a maximum for a particular value of $E_{0,blob2}$. By carefully adjusting the values of these parameters one can try to maximize the $f.o.m.$, and therefore the experimental sensitivity in the energy region under investigation.

The optimization process is highly demanding in computing resources: beyond being a multivariate problem, the determination of the $f.o.m.$ for every parameter configuration requires analyzing tens of thousands of events. We therefore adopted here a pragmatic approach, which allowed attaining a high value for the $f.o.m.$, but without proving that this is a global maximum. The analyzed dataset comprised $3.8\cdot10^4$ MC and $1.9\cdot10^4$ detector data events passing the energy, fiducialization and gross single-track cuts. We performed separate analyses to MC events using the true information (equations (\ref{eq:eps true})-(\ref{eq:fom true})), MC events where signal and background estimates were made using the fitting procedure (equations (\ref{eq: N_bkg,i})-(\ref{eq: bkg acceptance fit})), and experimental detector data using the same fitting procedure. The optimal choice of parameters and resultant $f.o.m.$ vary, to some extent, between MC and detector data. Appendix \ref{sec:RL_par} provides a detailed description of the optimization process. Here, we present the final results of the analysis.

Our choice of parameters aimed to simultaneously achieve a high $f.o.m.$ for both detector data and MC (using the fitting procedure) while looking for optimal MC (true information) performance. We refer to the chosen set of parameters as the ``optimal configuration,'' where: $q_{cut}=10$\;PE, $N_{iter}=75$, $\epsilon_{cut}=0.008$, $l_{voxel}=5$\;mm and $R_{blob}=18$\;mm.

\begin{figure}
	\begin{center}
		\includegraphics[width=1.0\textwidth]{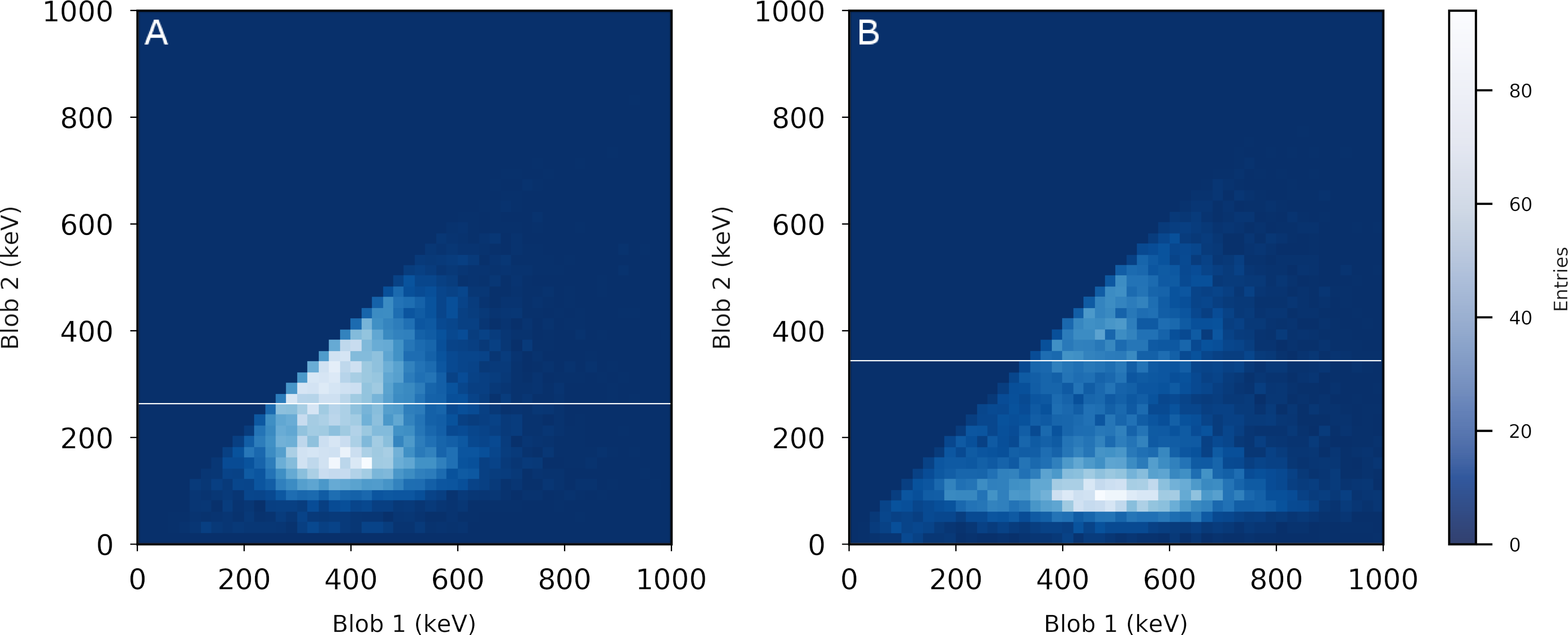}
		\caption{2D histogram of blob1 and blob2 energies of 1.6\;MeV double escape peak events from detector data: (A) classical analysis; (B) RL deconvolution with the optimal choice of parameters. Each histogram contains events passing the fiducial cut and gross single-track cut with energies in the range $1.4-1.8$\;MeV, with both signal and background populations (with higher and lower blob2 energies, respectively). The horizontal white lines mark the choice of blob2 energy cut which maximizes the $f.o.m$.}
		\label{img:blob1_2_2D_hist_DE_classical_vs_RL}
	\end{center}
\end{figure}

Figure \ref{img:blob1_2_2D_hist_DE_classical_vs_RL} shows 2D histograms of signal and background events from detector data, binned according to their blob1 and blob2 energies, comparing the classical analysis (A) to RL deconvolution with the optimal choice of parameters (B). Each histogram contains both signal (electron-positron) and background (Compton scatter) events, in the range $1.4-1.8$\;MeV, where the signal population occupies a higher region in the blob1-blob2 plane. The horizontal white lines mark the choice of blob2 energy cut which maximizes the $f.o.m$. Application of RL deconvolution clearly leads to an improved separation between the two groups.

\begin{figure}
	\begin{center}
		\includegraphics[trim={3.5cm 0.5cm 4cm 2cm}, clip, width=1.0\textwidth]{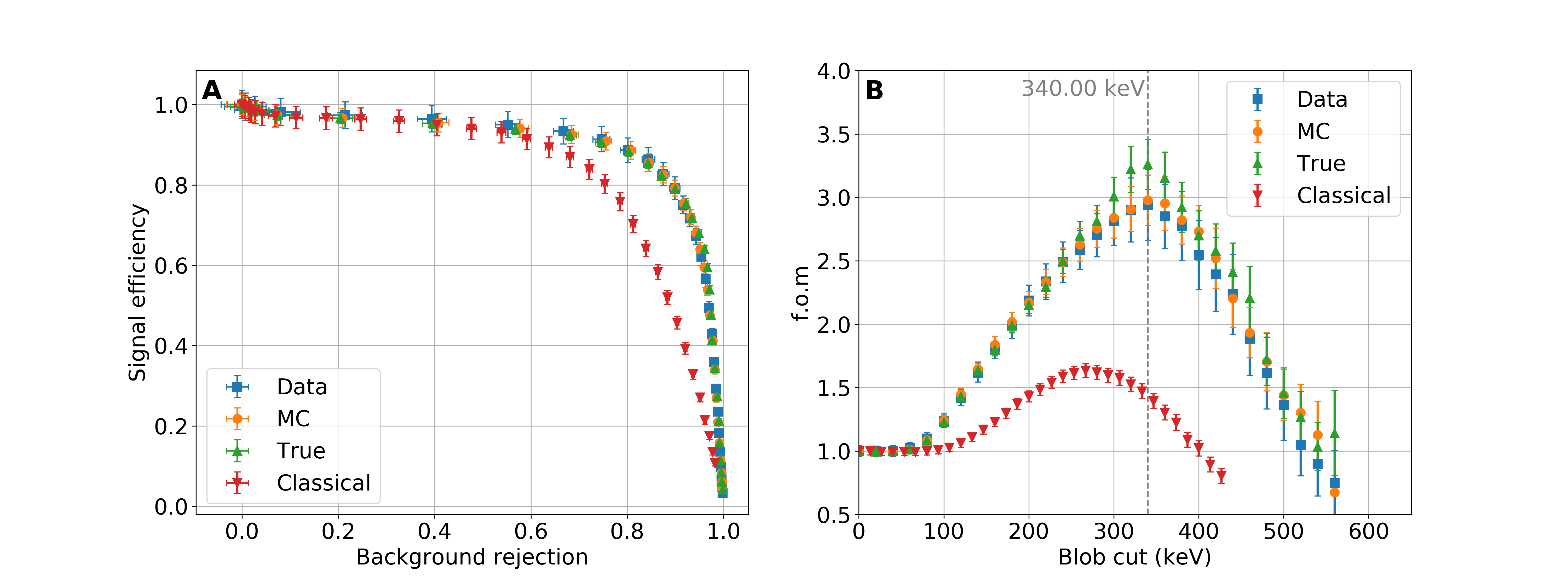}
		\caption{(A) Signal efficiency vs. background rejection for 1.6 MeV double escape peak events, for detector data, MC (using the fit) and MC (using the true information), for the optimal choice of parameter configuration (marked rectangles in figure \ref{img:parameter_scan} in appendix \ref{sec:RL_par}). The curve resulting from the classical analysis of data (from \cite{Ferrario:2019kwg}) is shown for comparison. (B) The figure of merit for the optimal parameter configuration for detector data, MC (fit and true) and classical analysis vs. the threshold on blob2 energy. The maximal (optimal) $f.o.m.$ is for a blob2 energy cut at 340 keV.}
		\label{img:pp signal efficiency and fom}
	\end{center}
\end{figure}

Figure \ref{img:pp signal efficiency and fom}A shows the signal efficiency as a function of background rejection ($1-b$) for 1.6 MeV double escape peak events, for the optimal choice of parameters. The figure includes the curves for detector data, MC (using the fit and true information) and -- for comparison -- the classical analysis of detector data events \cite{Ferrario:2019kwg}. Increasing background rejection is equivalent to moving the horizontal white lines in figure \ref{img:blob1_2_2D_hist_DE_classical_vs_RL} upward. Figure \ref{img:pp signal efficiency and fom}B shows the $f.o.m.$ as a function of the threshold on blob2 energy for the same datasets and parameters. For RL deconvolution the maximal (optimal) $f.o.m.$ is obtained for $E_{0,blob2}=340$\;keV. 

Table \ref{tab:table1} shows the signal efficiency, background acceptance and $f.o.m.$ for detector data and MC (with the classification based on both the fit and true information), for the optimal choice of parameters and optimal value of blob2 energy cut. For comparison the table also includes the results of the classical analysis. The errors represent statistical uncertainties (standard deviation); systematic effects were found to be dominated by the choice of reconstruction parameters and are not further considered here.

\begin{table}[!h]
	\begin{center}
		\caption{Signal efficiency and background acceptance for the optimal figure of merit.}
		\vspace{5mm}
		\label{tab:table1}
		\begin{tabular}{l|c|c|c}
			\textbf{Dataset/analysis} & \textbf{Signal efficiency} & \textbf{Background acceptance} & \textbf{Figure of Merit} \\
			\hline
			Classical & $71.6\pm1.5\%$ & $20.6\pm0.4\%$ &  $1.58\pm0.04$ \\
			Data      & $56.6\pm2.2\%$ & $3.7\pm0.7\%$  &  $2.94\pm0.28$ \\
			MC (fit)  & $59.4\pm1.6\%$ & $4.0\pm0.5\%$  &  $2.98\pm0.20$ \\
			MC (true) & $59.4\pm1.0\%$ & $3.3\pm0.4\%$  &  $3.26\pm0.20$ \\
			\hline
		\end{tabular}
	\end{center}
\end{table}

For detector data, the RL-based analysis using the optimal choice of parameters provides a 5.6-fold reduction of background acceptance compared to the classical analysis (overall topological background rejection factor of $\sim27$), accompanied by a relative reduction of signal efficiency by 21\%. According to the simulated data (true MC information), a 6.2-fold reduction of background and a 17\% relative reduction of signal is achieved with the chosen configuration.

\begin{figure}
	\centering
	\includegraphics[trim={3.0cm 0.5cm 4cm 2cm}, clip, width=1.0\textwidth]{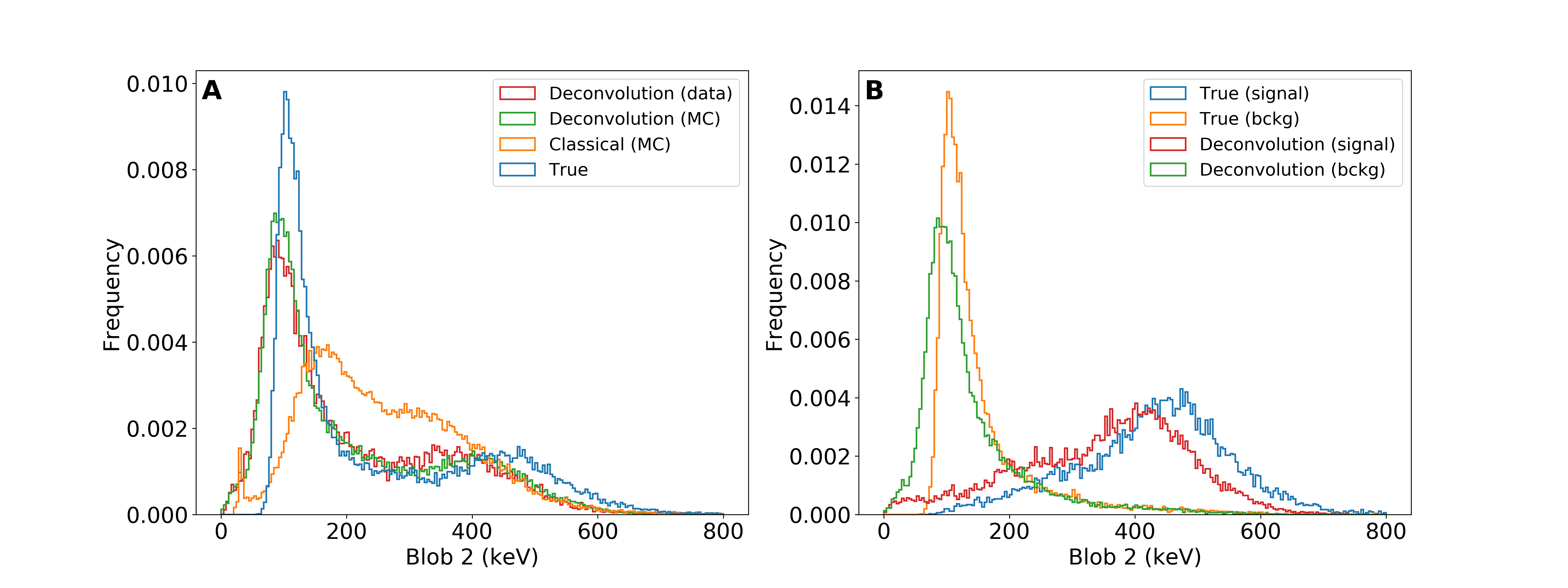}
	\caption{(A) Reconstructed energy for blob2 for deconvolution and classical analysis. The ``true'' blob energy is found by integrating over a sphere of 18\;mm radius centered on the true track end-point. (B) Distributions divided by population, either signal or background.}
	\label{img:blob_recoE}
\end{figure}

\begin{figure}
	\centering
	\includegraphics[trim={3.5cm 0.5cm 4cm 2cm}, clip, width=1.0\textwidth]{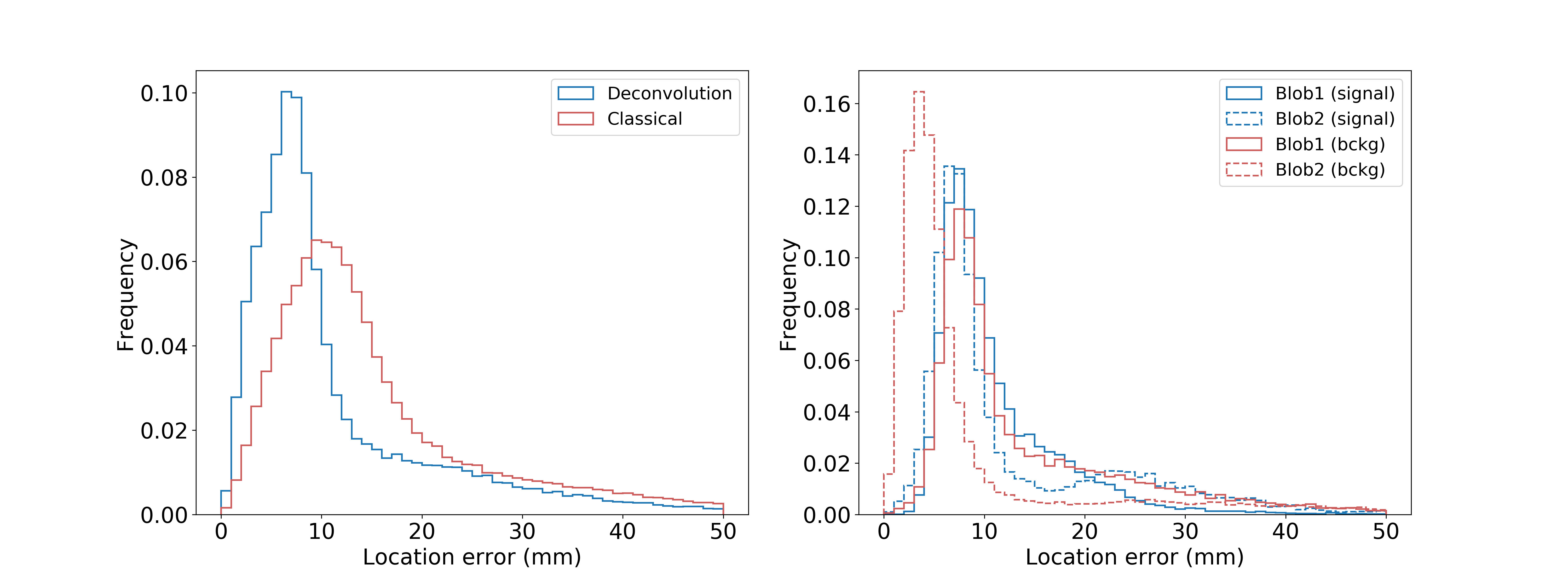}
	\caption{End-finding error distributions. Left: overall error distributions obtained by RL-deconvolution and the classical analysis for 1.6\;MeV double escape peak events; right: distributions for RL-deconvolution divided into signal and background events, with subdivision into blob1 and blob2.}
	\label{img:errorblob}
\end{figure}

The enhanced background rejection power is a consequence of the improved blob reconstruction, both in positioning and radial extent, enabled by RL deconvolution. The RL-reconstructed blob energy distribution is much closer to the expected one than that derived from the classical analysis, as illustrated in figure \ref{img:blob_recoE}. Only blob2 is shown as the difference in performance is more pronounced than for blob1 (which also exhibits an improvement). Figure \ref{img:errorblob}, left, compares the overall performance of RL deconvolution and the classical analysis in end-finding. The histograms show the distributions of distances between the reconstructed end positions and the true ones (extracted from MC) for all events in the 1.6\;MeV ROI, combining signal and background with no distinction between blob1 and blob2. Employing RL deconvolution clearly reduces the errors, leading to improved blob-based classification. In appendix \ref{sec:RL_par} we show that, as expected, high f.o.m. values are correlated with reduced errors in end-finding. The right panel of figure \ref{img:errorblob} focuses on the error distributions for RL-reconstructed events, separating them by event and blob type. It shows that the best results are obtained for blob2 in background events, where there is no Bragg peak. While there is considerable reduction in error for the other cases compared to the classical analysis, in about 30\% of the events the error is larger than the blob radius (18\;mm), indicating that there is still ample room for improvement. 

%% file: src/discussion.tex
\section{Discussion}

Richardson-Lucy deconvolution was shown in this work to be a highly effective tool for enhancing image reconstruction in NEXT, leading to a major improvement in topological background rejection. The application of the method to detector data in the 1.6\;MeV double escape peak of $^{208}$Tl, using a cut on blob2 energy, yielded a background acceptance level $b=(3.7\pm0.7)\%$ with a signal efficiency $\epsilon=(56.6\pm2.2)\%$ (figure of merit $f.o.m.=\epsilon/\sqrt{b}=2.94\pm0.28$), with similar results for detector data and simulated events. This represents a drastic improvement relative to previous results by the collaboration (20.6\% background acceptance with 71.6\% signal efficiency, $f.o.m.=1.58$ \cite{Ferrario:2019kwg}) and greatly boosts the background rejection realizable by the experiment. The obtained level of background acceptance is similar to the value previously reported by the Gotthard experiment using visual inspection of double escape peak events at 1.6\;MeV \cite{Gotthard1998}. The new results are also considerably better than those obtained using a deep convoluted neural network on non-deconvolved tracks to classify double escape peak events, where the background acceptance was 10\% and signal efficiency was 65\% ($f.o.m.=2.06$) \cite{Kekic:2020cne}.

The primary effect of employing RL deconvolution is the attainment of refined 3D track images, which allows better identification of the track ends, and therefore improved positioning of the blob centers and better estimates of the energy they contain. Improved end finding generally results from the enhanced resolving power offered by RL-deconvolution. Several illustrative examples for this are given in appendix \ref{sec:classification_examples}.

The focus of this work was on the analysis of experimental detector data recorded at the 1.6\;MeV $^{208}$Tl double escape peak. However, we also probed the possibility of implementing the RL-based method to MC signal and background events in the $Q_{\beta\beta}$ ROI, using the NEXT-White detector MC model. Preliminary analysis indicated that similar levels of background acceptance and signal efficiency are expected in this energy range. However, since NEXT-100 will operate under different conditions (15\;bar, 112\;cm maximal drift, 15.6\;mm SiPM spacing and a different EL and tracking plane geometry), such results are only of indicative nature, and a full simulation is deferred to a separate publication.

In spite of the improved event classification offered by the new method, the full potential Richardson-Lucy deconvolution is yet to be exploited by the collaboration, as the results presented in this work remain a first-approach evaluation using existing tools (e.g., the BFS algorithm), which may not be optimal for the fine-grained output of the RL procedure. In particular, although figure \ref{img:errorblob} shows a clear improvement in track-end finding, the distributions of deconvolved events have significant tails extending to large errors, requiring the use of large blob radii for the analysis. Presently, several ideas for potential improvement in event classification are under study. These include improved algorithms for end finding, as well as the potential use of Machine Learning approaches for the classification of high-definition reconstructed events. In addition, RL deconvolution can be further developed to a full 3D method instead of following the slice-by-slice approach employed in this work, which could lead to enhanced image quality along the drift direction. Lastly, image quality may greatly benefit from the use of low-diffusion gas mixtures, such as Xe-He \cite{Felkai2018, McDonald2019, Fernandes2020}, or Xe ``spiked'' with low concentrations of molecular additives \cite{Azevedo2016, Henriques2017, Henriques2019}. 

\begin{figure}
	\centering
	\includegraphics[width=1.0\textwidth]{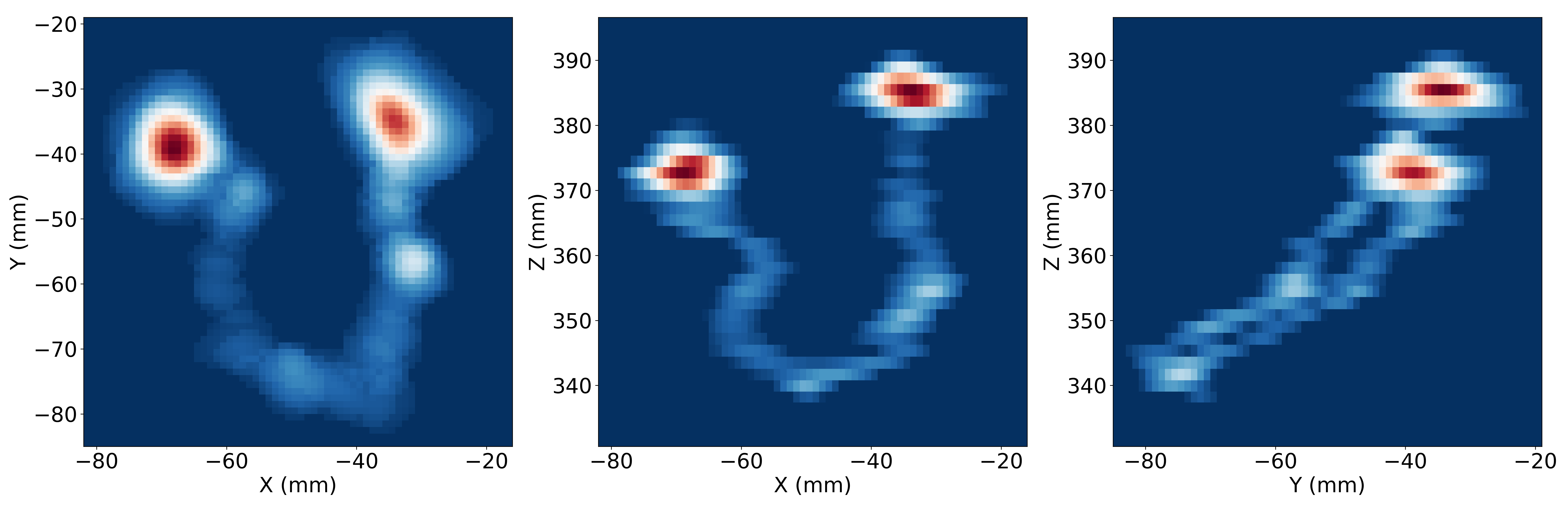}
	\caption{Deconvolved 2.0\;MeV $2\nu\beta\beta$ candidate obtained during the current data taking of NEXT-White.}
	\label{img:2nubb_track}
\end{figure}

Prior to further improvement of the method, the collaboration is presently evaluating the benefits offered by RL deconvolution in its current form, in particular for the analysis of $2\nu\beta\beta$ events in NEXT-White. As an example, figure \ref{img:2nubb_track} shows a 2.0\;MeV double beta candidate (from NEXT-White detector data) reconstructed by the RL procedure described above. 

%% file: src/RL_app.tex
\appendix

\section{Richardson-Lucy deconvolution} \label{sec:RL_app}

The Richardson-Lucy (RL) algorithm aims to recover, by deconvolution, an underlying sharp image from an observed blurred and noisy one. The algorithm is iterative, generating a sequence of improved approximations for the underlying image using the (presumably known) point spread function (PSF) of the imaging process. In this appendix we outline the mathematical procedure, as employed on 2D images. 

We denote by $W(x,y)$ the underlying sharp image and by $F(x,y)$ the PSF. In the absence of noise, the ideal blurred image $H(x,y)$ is obtained as a convolution of $W$ and $F$:

\begin{equation}
H(x,y)=\iint W(x',y') F(x-x',y-y') dx' dy' \label{eq: H ideal}
\end{equation}
In principle, $W(x,y)$ could be recovered from $H(x,y)$ by solving this integral equation. This could be done by discretization, converting $H$, $W$ and $F$ into matrices and the integral to a double summation. This generates a system of linear equations with the elements of $W$ as the unknowns:

\begin{equation}
\sum_{j}^{} F_{i,j}W_j = H_i \label{eq: linear equation system with H} 
\end{equation} 
where the indices $i$ and $j$ refer to individual elements in $H$ and $W$, combining the enumeration of both the row and column, and $F_{i,j}$ describes the influence of the $j$-th element of $W$ on the $i$-th element of $H$.
 
In reality, because of the presence of noise, the actual observed image $\tilde{H}(x,y)$ is different from the ideal one $H(x,y)$. The system of linear equations then becomes:

\begin{equation}
\sum_{j}^{} F_{i,j}W_j = \tilde{H}_i \label{eq: linear equation system with H tilde} 
\end{equation} 
Attempting to solve equations (\ref{eq: linear equation system with H tilde}) generally yields poor results, with large discontinuities in $\{W_j\}$, as well as non-physical negative values. This occurs because the process tends to amplify short-wavelength errors in $\tilde{H}$, which are characteristic of noisy images \cite{Lucy:1974yx}. 

The approach of the RL algorithm is different. It begins by noting that one could formally write:  

\begin{equation}
W(x,y)=\iint H(x',y') G(x-x',y-y') dx' dy' \label{eq: direct inversion}
\end{equation}
provided that we define the inverse kernel $G$ as:

\begin{equation}
G(x-x',y-y')=\frac{W(x,y) F(x-x',y-y')}{H(x',y')} \label{eq: G ideal}  
\end{equation}
where $\iint F(x-x',y-y') dx' dy' = 1$. Since $G$ depends on $W$, the direct calculation of $W$ from equation (\ref{eq: direct inversion}) is impossible. However, the process may work iteratively, if one could provide successively improved approximations for $G$, which, in turn, would rely on successive estimates of $W$. 

We begin with an initial estimate $W^{(0)}$ for $W$, where $W^{(0)}(x,y)$ is generally taken to be a flat image. In the $r$-th iteration we calculate an intermediate blurred image $H^{(r)}$ by:

\begin{equation}
H^{(r)}(x,y)=\iint W^{(r)}(x',y') F(x-x',y-y') dx' dy' \label{eq: H(r)}
\end{equation}   
This allows finding an estimate for $G$:

\begin{equation}
G^{(r)}(x-x',y-y')=\frac{W^{(r)}(x,y) F(x-x',y-y')}{H^{(r)}(x',y')} \label{eq: G(r)}  
\end{equation}

The new estimate for $W$, $W^{(r+1)}$, is then calculated following equation (\ref{eq: direct inversion}), with $\tilde{H}$ replacing $H$ and $G^{(r)}$ replacing $G$:

\begin{equation}
\begin{split}
W^{(r+1)}(x,y) &= \iint \tilde{H}(x',y') G^{(r)}(x-x',y-y') dx' dy' \\
&= W^{(r)}(x,y) \iint \frac{\tilde{H}(x',y')}{H^{(r)}(x',y')} F(x-x',y-y') dx' dy' \label{eq: W(r)}
\end{split}
\end{equation}

The discussion in \cite{Lucy:1974yx} shows that if successive changes in $W^{(r)}$ are sufficiently small, in the limit $r \rightarrow \infty$ the scheme converges to the solution of the corresponding maximum likelihood problem\footnote{The possibility of track reconstruction using a maximum likelihood approach had been previously explored by the NEXT Collaboration \cite{Simon:2017pck}, but was later disfavored for the method presented here.}. It further shows that if the number of elements in $H$ is equal or larger than those of $W$ (i.e., the system of linear equations (\ref{eq: linear equation system with H tilde}) is over-determined), this solution is unique.

\section{Parameter optimization} \label{sec:RL_par}

In this appendix we describe in detail the steps taken to optimize the choice of parameters used for the RL process and subsequent analysis.

\begin{figure}
	\centering
	\includegraphics[trim={0 0 0 0}, clip, height=6cm]{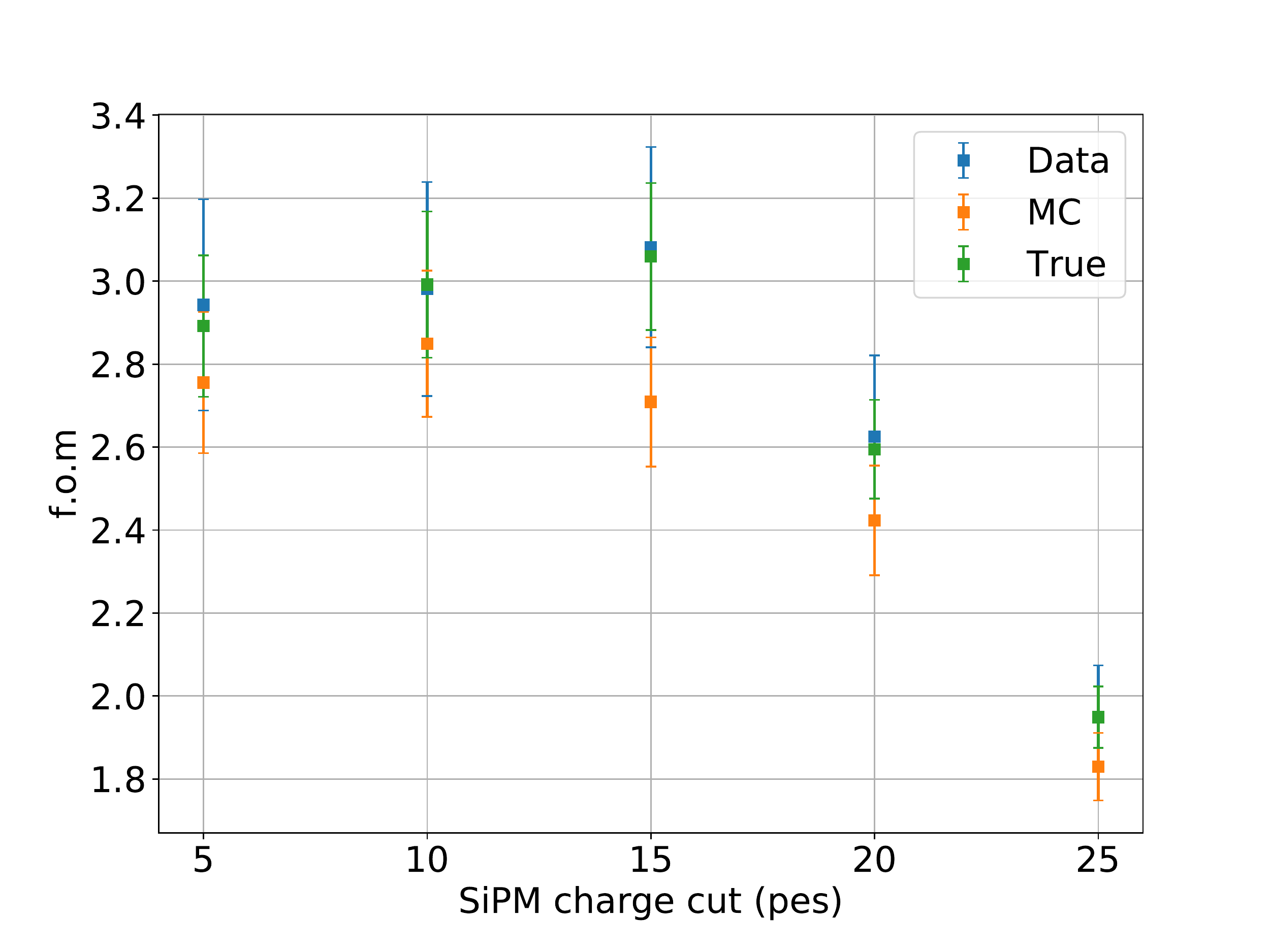}
	\caption{Variation of the maximal $f.o.m.$ achieved for both experimental detector data and MC for different cuts on the input signal. The scan on $q_{cut}$ was done with fixed values for the other parameters, as described in the text.}
	\label{img:qcut_fom}
\end{figure}

The first step in the analysis was to choose a value for the SiPM charge threshold, $q_{cut}$. It determines how much of the signal is cleaned out before starting the deconvolution process. Cutting too low may lead to the inclusion of distant signals (reflected light or photons induced by photoelectrons emitted from the gate), while cutting too high may bias and distort the output. The impact of $q_{cut}$ on the $f.o.m.$ was studied for several cut values, over the range $5-25$\;PEs (in 2 $\mu$s). We performed this scan keeping $N_{iter}=90$, $\epsilon_{cut}=0.008$ (in arbitrary units), $l_{voxel}=5$\;mm and $R_{blob}=21$\;mm (these values were chosen as a reasonable starting point after visual inspection of many events). For each value of $q_{cut}$ we calculated the $f.o.m.$ as a function of blob2 energy threshold and found its maximal value. Given the results, shown in figure \ref{img:qcut_fom}, we settled on a 10 PE cut due to a much better match between detector data and both MC fitted data and true information.

Next, we considered the effect of the number of RL iterations and final cleaning cut. Understanding the optimal point to stop applying RL iterations is of prime importance. If not applied enough times, the reconstructed charge distribution remains too blurry, which harms the blob energy estimation. On the other hand, over-iterating can result in noisy artifacts and in breaking up of the track to disconnected segments. These effects are strongly related to the subsequent application of the cleaning cut. If applied correctly, it can remove artifacts that appear in the iterating process. However, this is a delicate parameter as a too high cut could lead again into track fragmentation.

\begin{figure}
	\begin{center}
		\includegraphics[trim={13cm 1cm 12cm 3cm}, clip, width=1\textwidth]{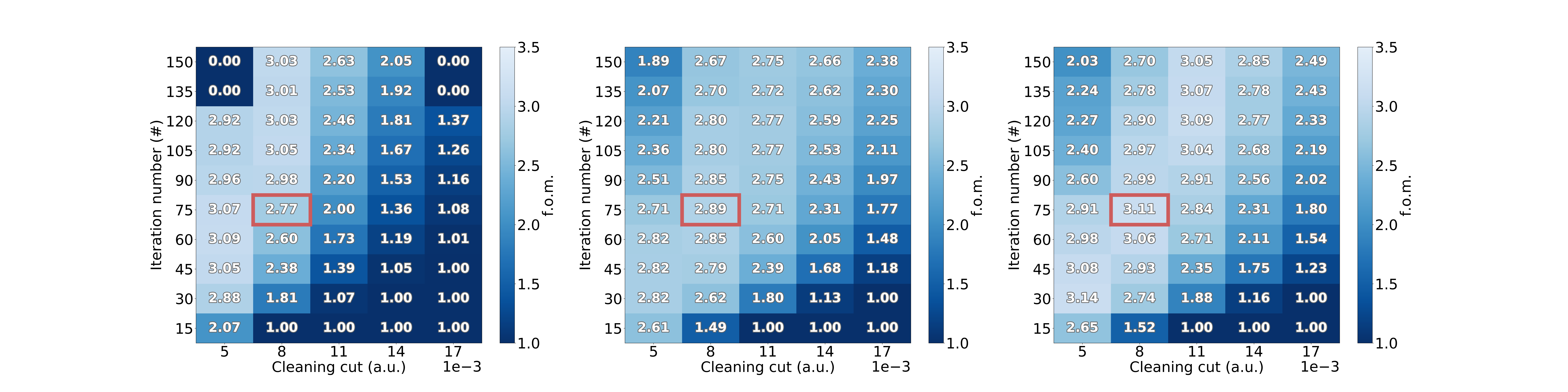}
		\put(-375, 113){Data}
		\put(-231, 113){MC}
		\put(-93, 113){True}
		
		\includegraphics[trim={13cm 1cm 12cm 3cm}, clip, width=1\textwidth]{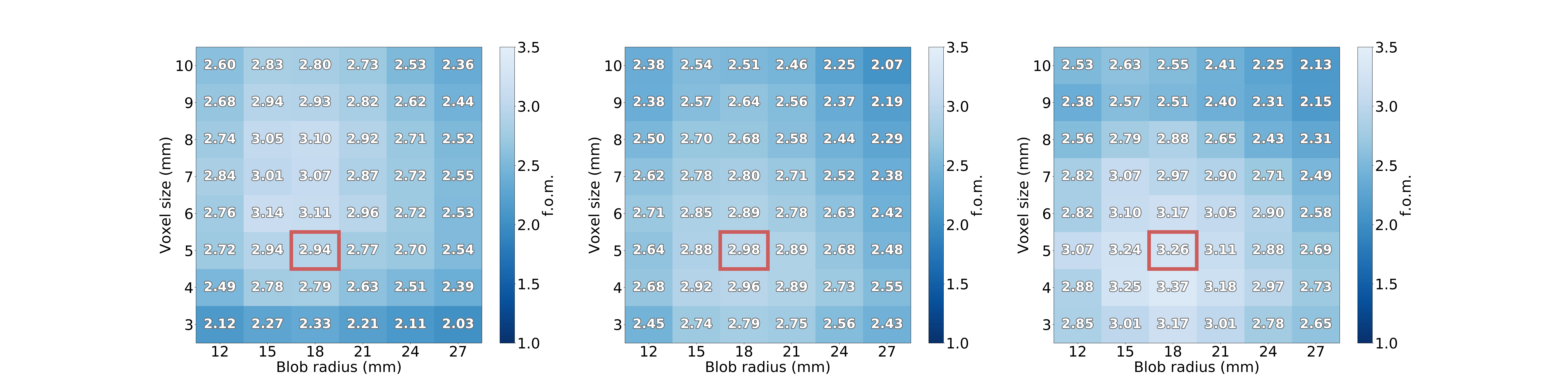}
		\caption{Parameter scan for optimizing the figure of merit, equation (\ref{eq:fom true}), for events in the double escape peak ROI for data, MC using the fitting procedure, and MC using the true information (i.e., whether the event contains a positron or not). The result for the optimal configuration is marked in red. Top row: maximal figure of merit for different combinations of the number of RL deconvolution iterations and the final threshold for the cleaning cut, keeping a voxel size of 5\;mm and blob radius of 21\;mm. Bottom row: maximal figure of merit for different combinations of voxel size and blob radius, for 75 RL iterations and a cleaning cut of 0.008. The SiPM charge cut is 10 PE in all cases.}
		\label{img:parameter_scan}
	\end{center}
\end{figure}

Given the observed relation between the number of iterations and the cleaning cut, we decided to scan and optimize both parameters simultaneously. The number of iterations was varied in steps of 15, and the cleaning cut was scanned over the range $0.005-0.017$. For this scan, we fixed $l_{voxel}=5$\;mm and $R_{blob}=21$\;mm. For each configuration we found the maximal $f.o.m.$ as a function of $E_{0,blob2}$. The results are shown in the top part of figure \ref{img:parameter_scan} for detector data, MC using the fit and MC using the true information. We chose a configuration which displayed a high $f.o.m.$ for both data and MC, including the true information, namely $N_{iter}=75$ and $\epsilon_{cut}=0.008$, yielding $f.o.m.(data) = 2.77$, $f.o.m.(MC\;fit) = 2.89$ and $f.o.m.(MC\;true) = 3.11$ (marked rectangles). 

With the above choice of $N_{iter}=75$ and $\epsilon_{cut}=0.008$ we moved to testing the effect of voxel size and blob radius. If voxels are too small, the track may be broken into disconnected segments, resulting in identifying internal points as the track ends. On the other hand, if voxels are too large, one loses the advantages of track refinement through RL deconvolution: the track is ``re-smeared'' and its ends are shifted. The blob radius is intimately related to the choice of voxel size. The blob must be large enough to contain the full energy deposited by the electron as it approaches the Bragg peak (which is shared between several voxels), but not too large, as this can lead to the inclusion of energy outside of the peak. Keeping $q_{cut}=10$\;PE, $N_{iter}=75$ and $\epsilon_{cut}=0.008$, we scanned $l_{voxel}$ from 3 to 10\;mm, and $R_{blob}$ from 12 to 27\;mm. As before, for each parameter configuration we scanned the threshold on blob2 energy and found the maximal $f.o.m.$ The results are shown in the second row of figure \ref{img:parameter_scan}, with rectangles marking the final choice of parameters, namely $l_{voxel}=5$\;mm and $R_{blob}=18$\;mm. We refer to this choice ($q_{cut}=10$\;PE, $N_{iter}=75$, $\epsilon_{cut}=0.008$, $l_{voxel}=5$\;mm and $R_{blob}=18$\;mm) as the ``optimal'' choice of parameters. 

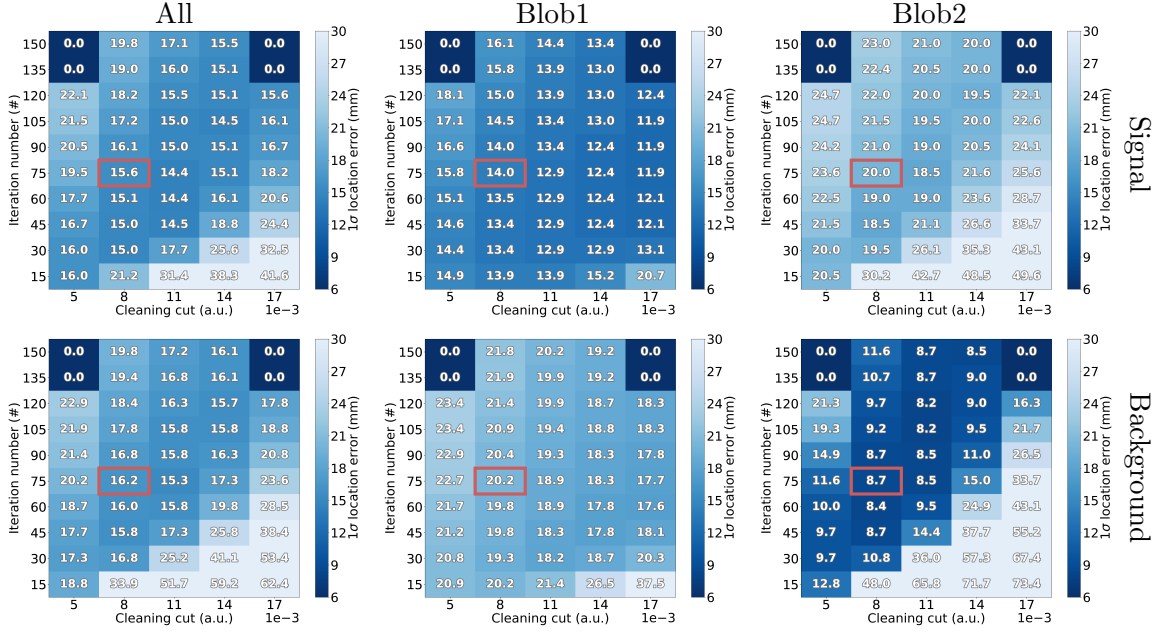
\begin{figure}
	\begin{flushleft}
		\includegraphics[trim={13cm 1cm 12cm 3cm}, clip, width=1\textwidth]{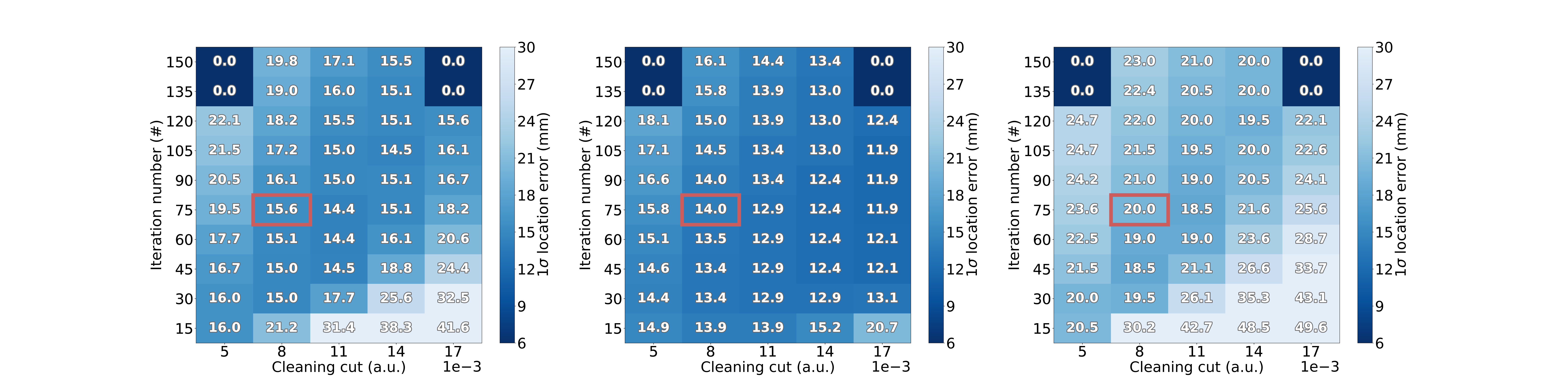}
		\put(-7,78){\rotatebox{270}{Signal}}
		\put(-370, 113){All}
		\put(-235, 113){Blob1}
		\put(-94, 113){Blob2}
		
		\includegraphics[trim={13cm 1cm 12cm 3cm}, clip, width=1\textwidth]{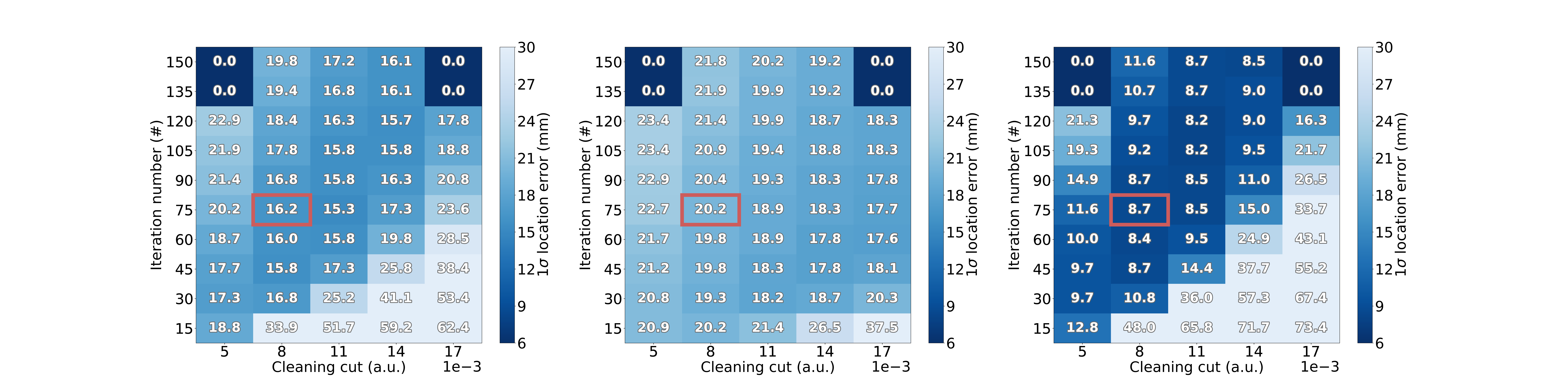}
		\put(-7,90){\rotatebox{270}{Background}}
		\caption{Blob location error distance for the RL parameter scan for MC events. The first column shows the overall value while the second and third columns show the blob1 and blob2 distribution respectively. The result for the optimal configuration is marked in red. Top row: signal events within the double escape peak ROI. Bottom row: background events in the same ROI.}
		\label{img:blob_parameter_scan}
	\end{flushleft}
\end{figure}

As a cross-check and validation of the choice of main RL parameters, namely the cleaning cut and number of iterations, the distance between the true location of the track ends and the reconstructed ones was computed and evaluated (for MC events), generating error distributions as shown in figure \ref{img:errorblob}. To quantify the analysis, we consider the error level below which the distribution contains 68\% of the events. We scanned the main RL parameters over a reasonable range of values, and calculated, for each configuration, the 68\% error level (to which we refer as the ``$1\sigma$ location error'').

\begin{figure}
	\begin{center}
		\includegraphics[trim={3.5cm 0.5cm 4cm 2cm}, clip, width=1.0\textwidth]{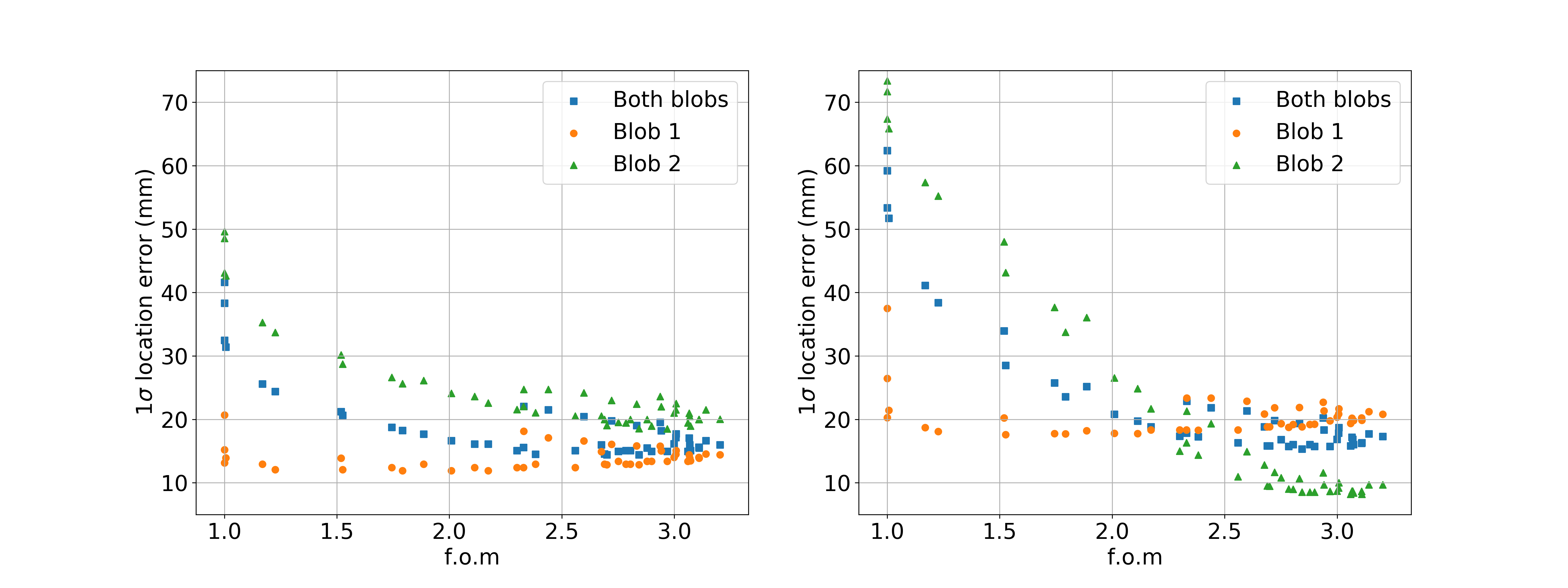}
		\put(-389,150){Signal}
		\put(-175,150){Background}
		\caption{Blob location error distance, defined in the text, for all configurations of the RL parameter scan and as a function of the best $f.o.m.$ value achieved with such configuration. Left: location error for signal events within the double escape peak ROI. Right: background events in the same ROI.}
		\label{img:fom_blob_distance}
	\end{center}
\end{figure}

The results of the scan are displayed on figure \ref{img:blob_parameter_scan}, which shows the $1\sigma$ location error obtained for different values of the number of iterations and cleaning cut for both blobs (first column), blob1 (second column) and blob2 (third column), where the scan results are shown separately for signal (first row) and background events (second row). Considering both blobs together (first column), the optimal choice of $N_{iter}$ and $\epsilon_{cut}$ (red rectangle) leads to an overall error close to the minimum, which would have been obtained for $N_{iter}=75$ and $\epsilon_{cut}=0.011$. Note that the blob1 population shows a much smaller change between configurations while the gradient in blob2 population is considerably stronger, especially in the case of background events. Figure \ref{img:fom_blob_distance} shows the end-location $1\sigma$ error values of figure \ref{img:blob_parameter_scan} as a function of the best $f.o.m.$ value obtained for each configuration using the true information (where the best $f.o.m.$ is calculated for the optimal choice of blob2 energy threshold). The data display a clear correlation between high $f.o.m.$ values and small errors in end-finding. Again, blob1 location error stays roughly constant while the blob2 location error improvement is fully correlated with increasing $f.o.m.$ values. This indicates that the overall improvement in the f.o.m. is primarily driven by the error in the location of blob2. 

\begin{figure}
	\begin{center}
		\includegraphics[trim={3.5cm 0.5cm 4cm 2cm}, clip, width=1.0\textwidth]{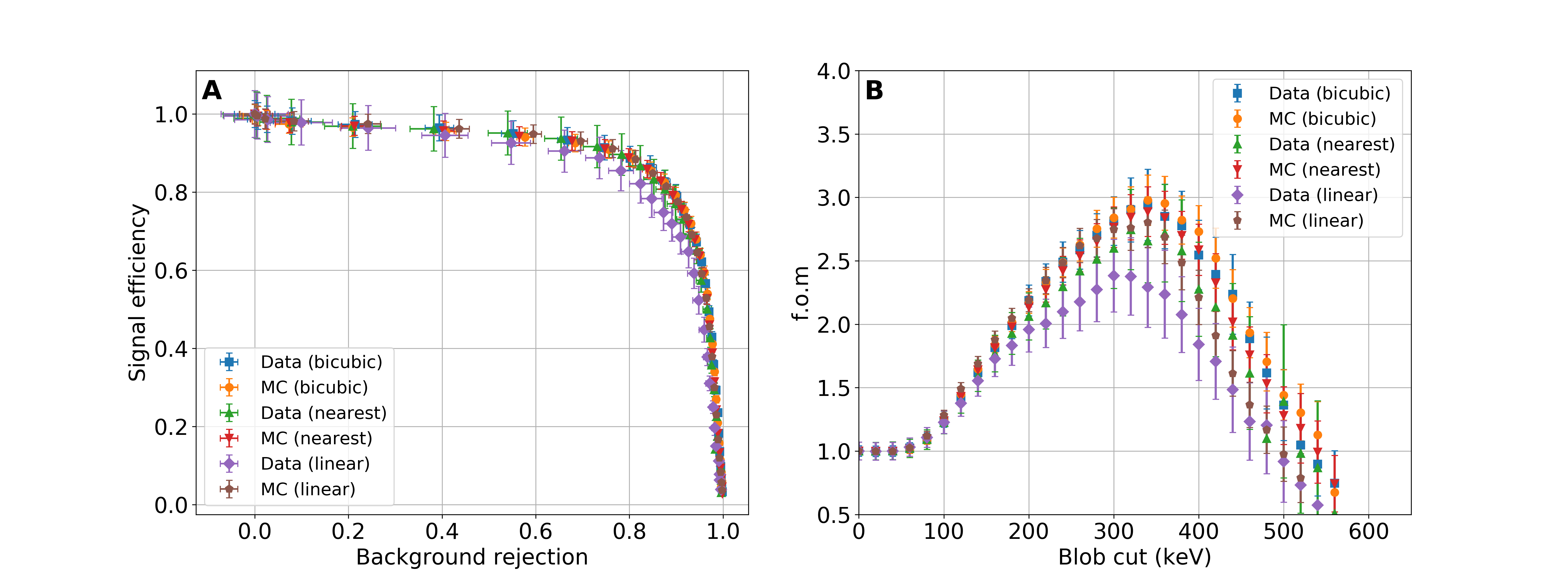}
		\caption{Comparing different interpolation approaches for NEXT-White data and MC (using the fit), for the optimal choice of RL parameters. (A) Signal efficiency vs. background rejection for 1.6 MeV double escape peak events. (B) The figure of merit for each dataset as a function of the blob2 cut.}
		\label{img:interpol_fom}
	\end{center}
\end{figure}

Lastly, a brief study of the impact of the interpolation method was performed. The evaluation consisted of repeating the topological analysis using the optimal configuration but with different interpolation approaches, to compare the figure of merit. In addition to the bicubic interpolation, we evaluated a linear interpolation and a nearest-neighbor approach, were the value assigned to each $10\times10$\;mm$^2$ bin is equal to the number of photons detected in the closest SiPM. The results are shown in figure \ref{img:interpol_fom}. While mostly all configurations are compatible within error, the maximal performance is achieved with the bicubic interpolation. This was expected as the parameters were optimized using bicubic interpolation and it is possible that different results can be achieved with the other approaches if optimizing the parameters for those. However, this study is left for future work.

\section{Examples of successful RL-based classification vs. failure of the classical analysis} \label{sec:classification_examples}

\begin{figure}
	\centering
	\includegraphics[trim={5cm 1cm 5.5cm 3cm}, clip, width=1.0\textwidth]{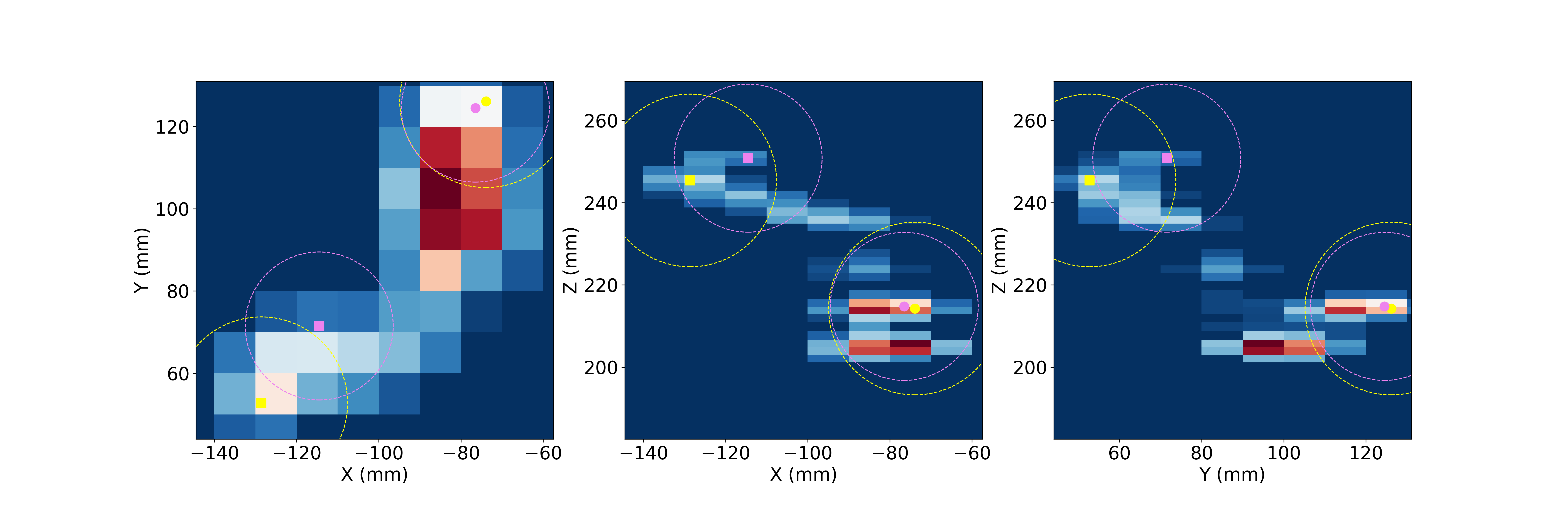}
	\put(-2, 131.5){\rotatebox{270}{Event 1: binned SiPM}}
	
	\includegraphics[trim={5cm 1cm 5.5cm 3cm}, clip, width=1.0\textwidth]{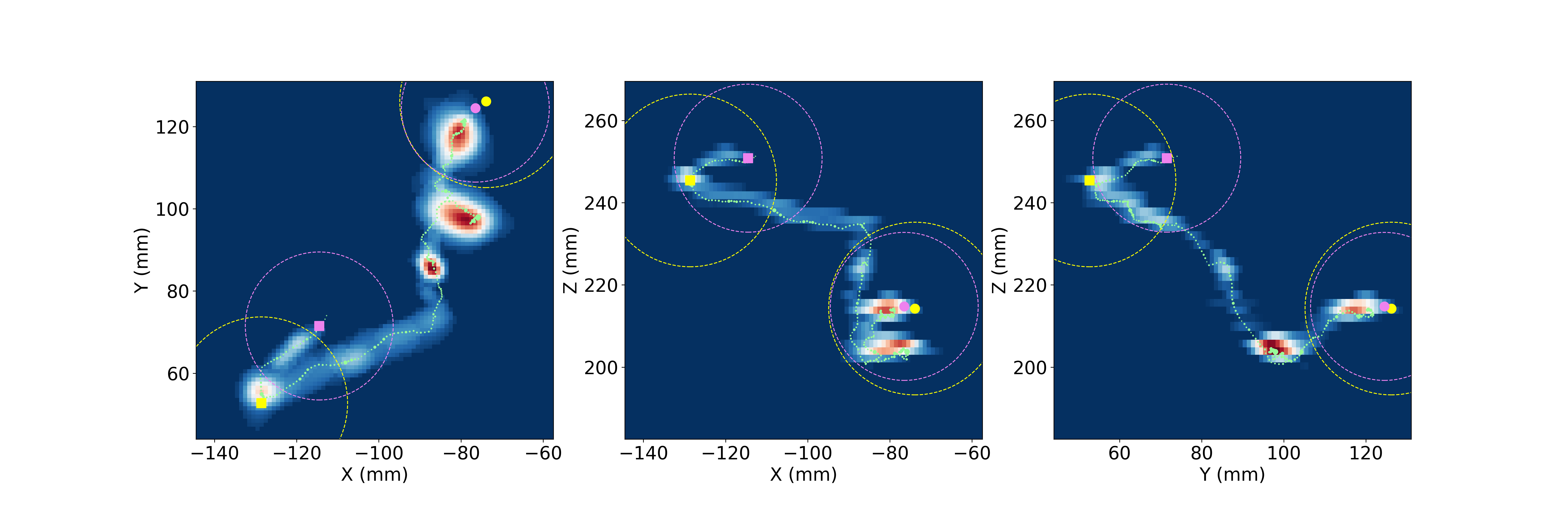}
	\put(-2, 133.5){\rotatebox{270}{Event 1: deconvolution}}
	
	\includegraphics[trim={5cm 1cm 5.5cm 3cm}, clip, width=1.0\textwidth]{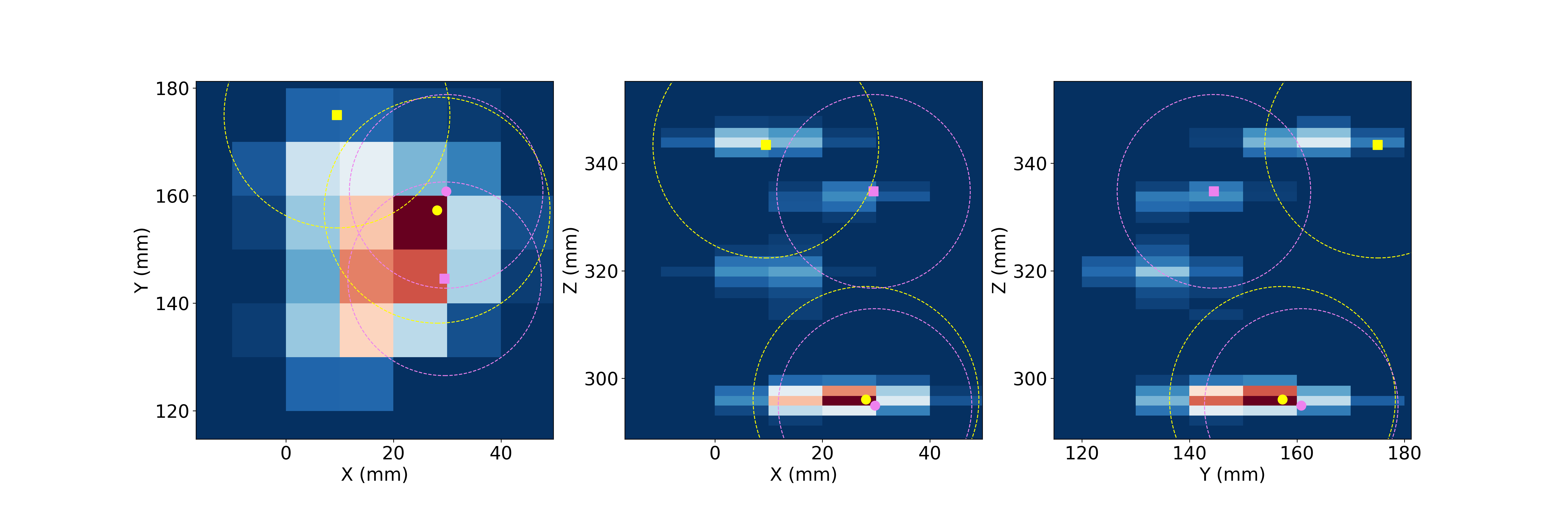}
	\put(-2, 131.5){\rotatebox{270}{Event 2: binned SiPM}}
	
	\includegraphics[trim={5cm 1cm 5.5cm 3cm}, clip, width=1.0\textwidth]{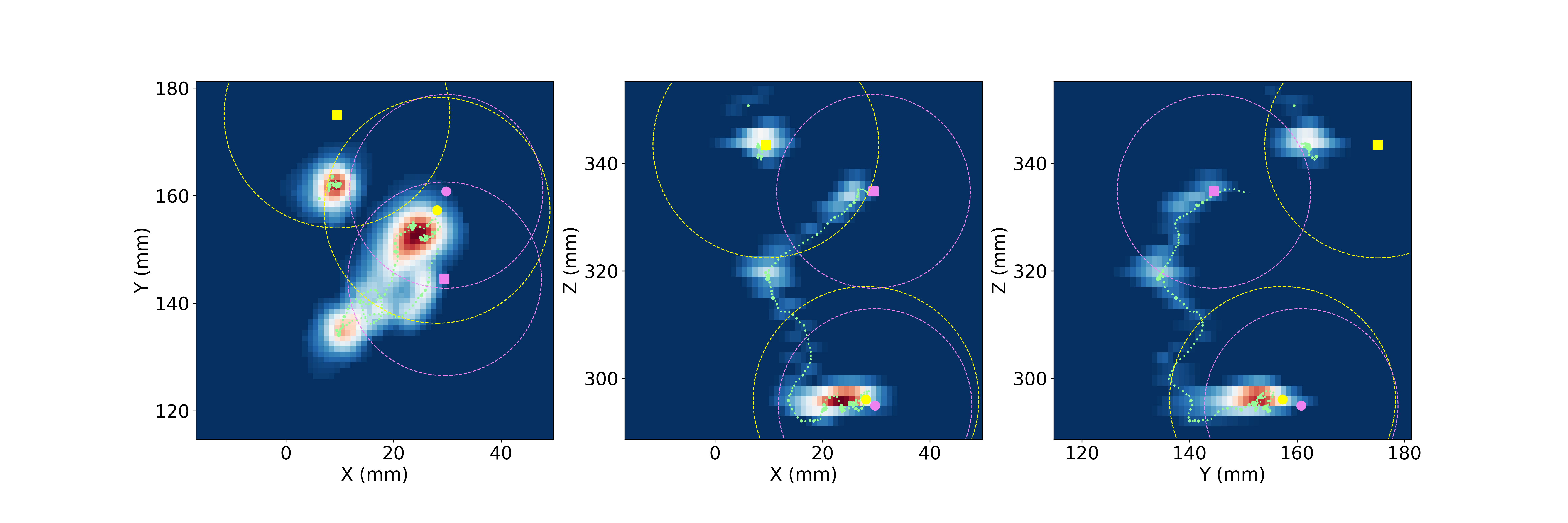}
	\put(-2, 133.5){\rotatebox{270}{Event 2: deconvolution}}
	
	\caption{Single electron events (background) in the pair production ROI misclassified by the classical analysis as signal and correctly identified as background by the RL process. For each event the top row is the binned SiPM response, and the bottom row -- the deconvolved projections. Yellow symbols mark the track ends found by the classical analysis, purple -- by the RL process. Squares are blob2 centers, circles -- those of blob1. The true track is overlaid on the deconvolved images. Misclassification of Event\;1 results from a merger between the start point of the track and its main part. Event\;2 (double Compton scatter) is misidentified as a single track because of the limited resolving power of the classical analysis.}
	\label{img:Better_end_finding_merger_and_single_track_errors}
\end{figure}

\begin{figure}
	\centering
	\includegraphics[trim={5cm 1cm 5.5cm 3cm}, clip, width=1.0\textwidth]{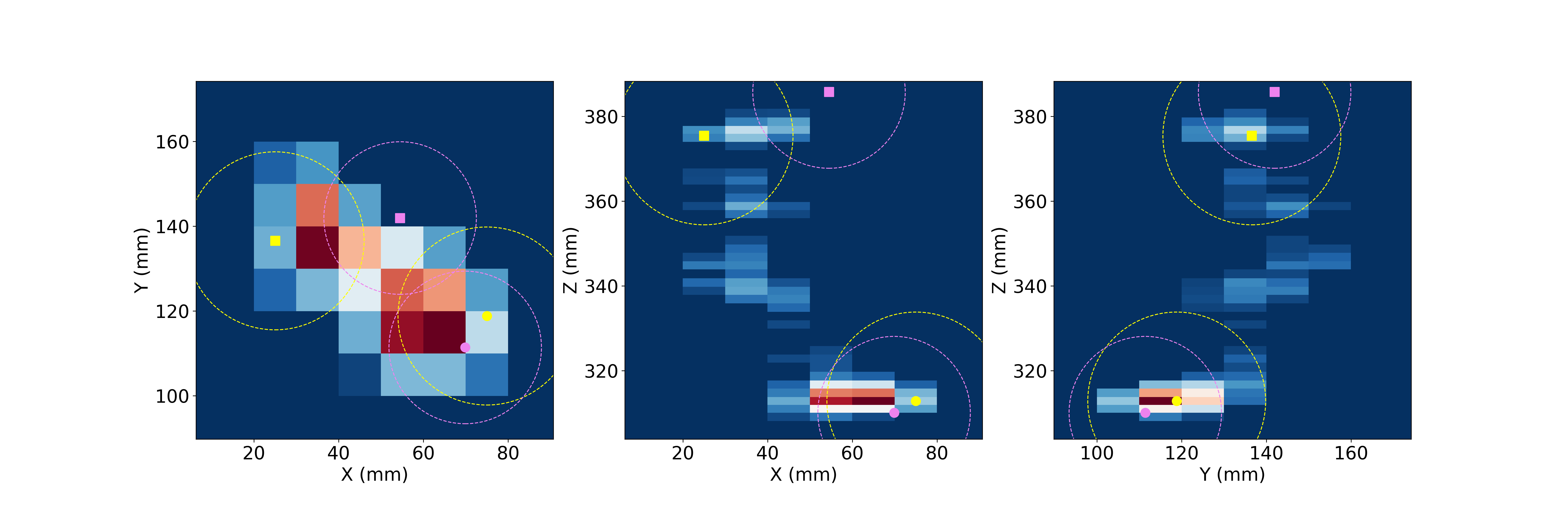}
	\put(-2, 131.5){\rotatebox{270}{Event 1: binned SiPM}}
	
	\includegraphics[trim={5cm 1cm 5.5cm 3cm}, clip, width=1.0\textwidth]{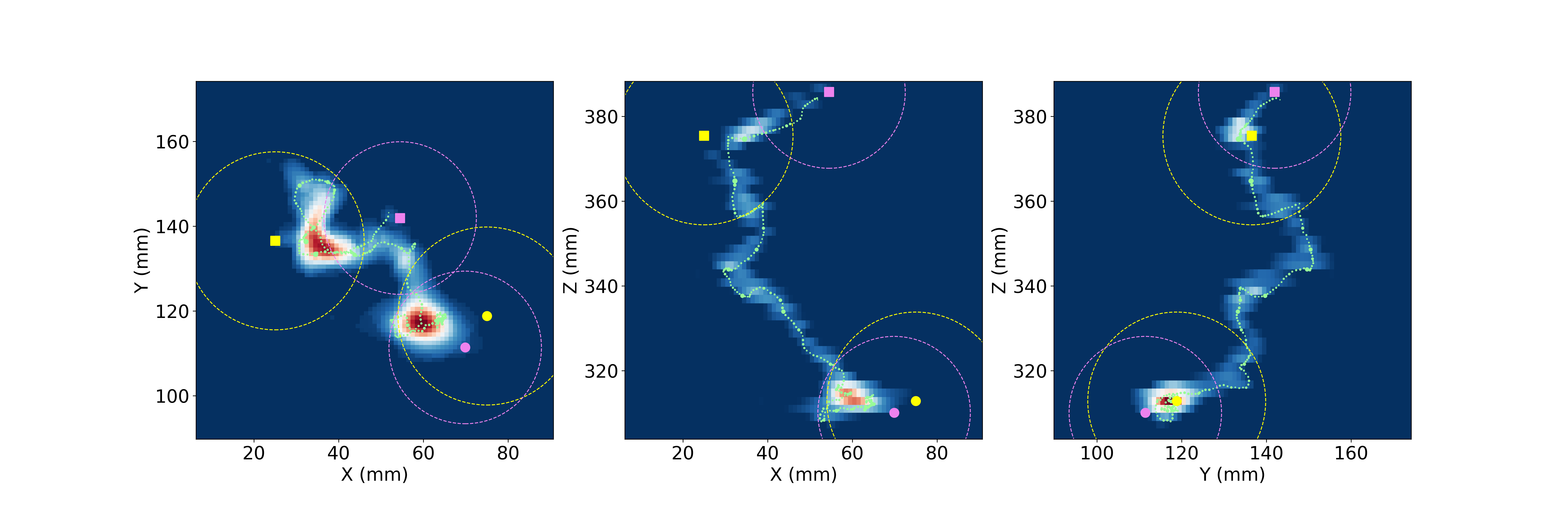}
	\put(-2, 133.5){\rotatebox{270}{Event 1: deconvolution}}
	
	\caption{An additional background event misidentified as signal by the classical analysis and correctly identified as background by the RL process. Symbols have the same meaning as in figure \ref{img:Better_end_finding_merger_and_single_track_errors}. Here the mistaken classification results from the high charge cut used in the classical analysis which removes $\sim2$\;cm of the start of the track.}
	\label{img:Better_end_finding_high_qcut}
\end{figure}

In this appendix we discuss three representative examples illustrating the reasons for improved classification of signal and background events by the RL-based method. The examples shown in figures \ref{img:Better_end_finding_merger_and_single_track_errors} and \ref{img:Better_end_finding_high_qcut} are simulated MC background events in the 1.4-1.8\;MeV ROI, which are misclassified by the classical analysis as signal, and correctly identified as background by the RL process. For each event, the top row shows the three projections of the raw sensor response, binned in $10\times 10\times 1.8$\;mm$^3$ voxels, with a charge cut of 30\;PE for the classical analysis. The bottom row shows the corresponding projections after RL deconvolution (charge cut of 10\;PE with 75 iterations). Yellow symbols designate the track ends found by the classical analysis and purple symbols - the ones found by the RL process. Squares represent the center of blob2 and circles those of blob1. The deconvolved images further include the true track overlaid in green. The circles represent the blobs used in both analysis methods.

Event\;1 is a photoelectric absorption of a Compton scattered gamma, accompanied with a delta electron, where the track is ``folded'' such that its true start point is close to its main part. Since the classical analysis rebins the SiPM hits in 15\;mm voxels, the true start point (purple square) is merged into the track ``body'', and a distant internal point (yellow square) is misidentified as the track extremity. Since the local ionization density near this point is high, the energy contained in the classical blob2 centered at it lies above the threshold value, and the event is identified as signal. The RL process, on the other hand, identifies both ends with an error of a few mm, and correctly places the center of blob2 close to the track starting point.

Event\;2 is a double Compton scatter of a $^{208}$Tl 2615\;keV gamma. The first scatter (starting point of the main track, close to the purple square) gives rise to an energetic Compton electron that creates the main track. The scattered photon interacts $\sim2$\;cm from the first vertex, generating a Compton electron of lower energy and a low-energy gamma, which interacts at a short distance. The classical analysis fails to separate the two Compton electron tracks and places the center of blob2 close to the Bragg peak of the second electron. Since the energy contained in this blob is above threshold the event is identified as signal. RL deconvolution, in contrast, successfully resolves the two tracks and places blob2 center close to the first Compton vertex.

Event\;3 (figure \ref{img:Better_end_finding_high_qcut}) provides an example for the effect of reduced SiPM charge threshold ($q_{cut}$) in the RL process. The 30\;PE cut employed in the classical analysis (after careful optimization, as discussed in \cite{Ferrario:2019kwg}) eliminates $\sim2$\;cm of the low ionization-density tail of this event. By chance, the classical blob2 energy lies above threshold, which leads to the misidentification of this event as signal. The same track, under the RL process, is fully reconstructed in the tail region, with proper placement of blob2 energy at the true starting point of the photoelectron.